\documentclass[final,5p,times,twocolumn]{elsarticle}

\usepackage{ecrc}
\usepackage{amsthm}
\usepackage{mathtools}
\usepackage{amssymb}
\usepackage{xcolor}
\usepackage{color} 
\usepackage{graphics}
\usepackage{placeins}
\usepackage{bm}

\volume{00}

\firstpage{1}

\journalname{Communications in Nonlinear Science and Numerical Simulation}
\journal{Communications in Nonlinear Science and Numerical Simulation}
\runauth{B. Many Manda et al.}

\jid{CNSNS}

\jnltitlelogo{CNSNS}

\CopyrightLine{2021}{Published by Elsevier Ltd.}

\usepackage{amssymb}

\usepackage[figuresright]{rotating}

\begin{document}

\begin{frontmatter}



\dochead{}

\title{
Efficient detection of chaos through the computation of the Generalized Alignment Index (GALI) by the multi-particle method
}


\author[1,2]{Bertin Many Manda}
\author[1,3,4]{Malcolm Hillebrand}
\author[1]{Charalampos Skokos}
\address[1]{Nonlinear Dynamics and Chaos Group, Department of Mathematics and Applied Mathematics, University of Cape Town, Rondebosch, 7701 Cape Town, South Africa}
\address[2]{Laboratoire d’Acoustique de l’Universit\'e du Mans (LAUM), UMR 6613, Institut d’Acoustique - Graduate School (IA-GS), CNRS, Le Mans Universit\'e, France}
\address[3]{Max Planck Institute for the Physics of Complex Systems, N\"othnitzer Stra\ss e 38, 01187 Dresden, Germany}
\address[4]{Center for Systems Biology Dresden, Pfotenhauer Stra\ss e 108, 01307 Dresden, Germany}

\begin{abstract}
    We present a method for the computation of the Generalized Alignment Index (GALI), a fast and effective chaos indicator, using a multi-particle approach that avoids variational equations. We show that this approach is robust and accurate by deriving a leading-order error estimation for both the variational (VM) and the multi-particle (MPM) methods, which we validate by performing extensive numerical simulations on two prototypical models: the two degrees of freedom H\'enon-Heiles system and the multidimensional $\beta$-Fermi-Pasta-Ulam-Tsingou chain of oscillators. The dependence of the accuracy of the GALI on control parameters such as the renormalization time, the integration time step and the deviation vector size is studied in detail. We test the MPM implemented with double precision accuracy ($\varepsilon \approx 10^{-16}$) and find that it performs reliably for deviation vector sizes $d_0\approx \varepsilon^{1/2}$, renormalization times $\tau \lesssim 1$, and relative energy errors $E_r \lesssim \varepsilon^{1/2}$.     These results hold for systems with many degrees of freedom and demonstrate that the MPM is a robust and efficient method for studying the chaotic dynamics of Hamiltonian systems.    Our work makes it possible to explore chaotic dynamics with the GALI in a vast number of systems by eliminating the need for variational equations.
\end{abstract}




\end{frontmatter}


\section{Introduction}
\label{sec:intro}

Revealing the  properties of individual or ensembles of orbits is of fundamental interest for nonlinear systems. A key set of tools for this task are {\it chaos indicators}, real-valued quantities which exhibit  distinct behaviors for chaotic and regular dynamics~\cite{H2000,CARPINTERO201419,SGL2016,PP2016}. Preeminent among these indicators are the Lyapunov exponents (LEs)~\cite{BGGS1980a,BGGS1980b} which quantify chaos~\cite{S2010,PP2016,AKEDBK2018,RG2020} and can indicate the existence of first integrals of the dynamics~\cite{S2024,SM2024} in nonlinear systems, based on the growth rate of small deviations to a given trajectory. Following the LEs, the importance of detecting chaos has led to a plethora of new techniques.These include the fast Lyapunov indicator (FLI)~\cite{FLG1997,LGF2016}, the mean exponential growth of nearby orbits (MEGNO)~\cite{CS2000,CG2016} the smaller (SALI) and generalized (GALI) alignment indices~\cite{S2001,SBA2007,SM2016}, which have found applications in a broad array of fields such as information theory~\cite{LSB2019}, biophysics and materials~\cite{HKSS2019}, astrophysics \cite{ZDSS2020,KES2022,HHH2022}, as well as population dynamics and epidemiology \cite{G2021,SAZ2021,BRM2024}.

The wide-ranging importance of these indicators has motivated extensive work on how to effectively and efficiently detect chaos, whether through optimizing the numerical procedures \cite{SG2010,GES2012,SS2018}, choosing appropriate methods of calculation~\cite{TSR2001,MH2018}, or using short-time computations to determine chaoticity~\cite{HZNKWS2022}. The GALI method has been particularly successful in reliably and rapidly identifying chaos~\cite{SBA2007,BGM2023}, and is as such worthy of investment into finding ways of optimizing and generalizing its applicability in nonlinear dynamics.

These chaos indicators, including the GALI, rely on the time evolution of  deviation (tangent) vectors to a trajectory under investigation. One approach for obtaining these deviation vectors is to generate a reference trajectory along with trajectories starting nearby. Then the deviation vectors are calculated directly by estimating the separation vectors between the nearby trajectories and the reference one -- an approach we term the multi-particle method (MPM). Depending on whether the neighboring orbits are kept in the vicinity of, or are driven away from, the system's reference orbit allows the characterization of the studied trajectory's long-time behavior.  On the other hand, since the nearby trajectories have to be infinitesimally close to the reference one in the phase space, one can generate the dynamics of these deviation vectors by linearizing the system's equations of motion around the studied trajectory, yielding the so-called {\it variational equations}. The set of the equations of motion and the variational equations can therefore be integrated directly. 
This method is commonly known as the variational method (VM).

Due to its accuracy and efficiency, the VM is typically used for chaos detection~\cite{DK2018} and as such has been the preferred approach in the definition and construction of recent chaos indicators. Nevertheless, the VM requires the analytical expression of the equations of motion which must be continuous and differentiable over the considered space-time domain. Such strong requirements cannot always be satisfied, for example in systems in which elements interact following Hertzian contacts~\cite{J1987}, non-analytic softening/hardening laws~\cite{LWKH2008,CPGC2009} and in systems defined by empirical fields as in complex chemical reactions~\cite{Schlegel2003,SW2018,Naidoo2021}. Further, even in cases where it is possible for the variational equations to be derived, as models become more complex these equations can be impractical and computationally inefficient to work with (see e.g.~\cite{HMKGS2020}).

We note that the MPM framework has been proposed to address the shortcomings of the VM for the computation of the  LEs. In this context, several studies on the reliability of the MPM for the calculation of LEs have already been carried out~\cite{TSR2001,GES2012,MH2018} (see also Sect.~3.5 of~\cite{PP2016}). In these works it was found that the computation of the LEs can be affected by factors like the size of the initial separation vectors or the renormalization procedures.
It is worth noting that arbitrary choices of control parameters for the computation of LEs using the MPM can lead to spurious estimates as observed for example  in false positive MLE estimations in planetary motion~\cite{TSR2001}.
These results raise the question of how other chaos indices perform within the MPM framework, an area that remains largely unexplored. In this study, we focus on the GALI method, seeking to resolve several unanswered questions regarding its computation in the context of MPM. For instance, can the GALI be effectively defined within the MPM framework, and how does the evaluated quantity compare to its computation using the VM? How are GALI computations influenced by the MPM? Finally, can GALI, when computed via the MPM, accurately and efficiently detect chaotic behavior?

The aim of our study is to investigate the reliability of the GALI computation through the MPM. To achieve this goal we introduce a definition for the GALI based on the MPM formulation and demonstrate that it is equivalent to using the VM. We then obtain leading order analytical expressions of the numerical accuracy during the computation of the GALI values for both methods, and validate our analytical results by simulating the short-time dynamics of some well-known, prototypical nonlinear Hamiltonian systems: the H\'enon-Heiles (HH)~\cite{HH1964} and $\beta$-Fermi-Pasta-Ulam-Tsingou ($\beta$-FPUT)~\cite{FORD1992,BI2005} models.

We find optimal numerical parameters for computing the GALI with the MPM, and show that this computation enables full-scale nonlinear dynamics investigations at the same accuracy level as using the VM. We also examine the ability of the GALI method to reveal the global dynamics of Hamiltonian systems, once again probing the HH system, recapitulating known results and providing quantitative insights into the chaotic dynamics. These findings show that with well-chosen parameters it is entirely possible to use the MPM to accurately calculate the GALI, opening the way for a large variety of future studies of chaotic dynamics in complex systems.

The content of the paper is organized as follows.
Section~\ref{sec:model} contains a brief introduction to the LEs and the GALI. In Sect.~\ref{sec:theoretical_estimate}, we present the theoretical analysis for the accumulation of numerical errors in the computation of the GALI. In Sect.~\ref{sec:numerical_results}, we examine the conditions of reliability for the computation of the GALI using both the VM and MPM, focusing on regular and chaotic orbits of the HH and the $\beta$-FPUT models.  Finally in Sect.~\ref{sec:conclusion_and_outlook}, we summarize our findings and discuss some open questions. 
In the appendices, we provide a number of details relevant to this work, including a pseudo-code for the computation of GALI using the MPM, initial conditions for some orbits considered in the main body of the paper, and additional GALI results for the $\beta$-FPUT model.

\section{Lyapunov exponents and the generalized alignment index}
\label{sec:model}

In our study we consider autonomous Hamiltonian systems of $N$ degrees of freedom ($N$D), which are defined by a Hamiltonian function (whose numerical value is typically called the system's energy) of the form
\begin{equation}
  \mathcal{H}\left(q_1, q_2, \ldots, q_N, p_1, p_2, \ldots, p_N\right) = \mathcal{H}\left(\bm{x}\right),
  \label{eq:hamilton_general}
\end{equation}
where  $q_i$ and $p_i$, $i=1, 2, \ldots, N$ respectively are the generalized canonical coordinates and momenta of the system and $\bm{x} = \left(x_1, \allowbreak x_2, \allowbreak \ldots, x_{N},\allowbreak x_{N+1}, \ldots,\allowbreak x_{2N}\right)= \allowbreak\left(q_1,\allowbreak q_2, \allowbreak\ldots, \allowbreak q_N,\allowbreak p_1, \allowbreak \ldots, \allowbreak p_N\right)$ is a state vector in the system's phase space. The equations of motion are given by  
\begin{equation}
  \dot{\bm{x}} = \bm{J}_{2N}\cdot \frac{\partial \mathcal{H}}{\partial \bm{x}} = \bm{f}\left(\bm{x}\right),
  \label{eq:equation_of_motion_general}
\end{equation}
with 
\begin{equation}
  \bm{J}_{2N} = 
  \begin{pmatrix}
    \bm{0}_N & \bm{I}_N \\
    -\bm{I}_N & \bm{0}_N 
  \end{pmatrix},
  \label{eq:symplectic_matrix}
\end{equation}
being the so-called  symplectic matrix, where $\bm{0}_N$ and $\bm{I}_N$ are the $N\times N$ null and identity matrices respectively. In Eq.~\eqref{eq:equation_of_motion_general}, the over-dot denotes the derivative with respect to time $t$. Considering an initial condition $\bm{x}_0$ at time $t_0$, the solution of Eq.~\eqref{eq:equation_of_motion_general}, $\bm{x}(t)$, $t>0$, conserves the value of the system's energy $\mathcal{H}$ [Eq.~\eqref{eq:hamilton_general}]. In addition, we set $t_0 = 0$.

An $N$D Hamiltonian system has $2N$ (possibly non-distinct) LEs, ordered  as $\Lambda_1 \geq \Lambda_2 \geq \cdots \geq
\Lambda_{2N}$ . 
In the case of autonomous Hamiltonian systems, these exponents come in pairs of opposite signs, implying that $\sum_{i=1}^{2N} \Lambda_i= 0$.
In addition, due to the presence of an integral of motion (the Hamiltonian function itself), at least two of the LEs are always equal to zero~\cite{BGGS1980a,BGGS1980b,S2010,PP2016,SPM2020}.
The most commonly used chaos indicator is the estimation of the maximum LE (MLE) $\Lambda_1$, as  $\Lambda_1>0$ indicates chaos, while $\Lambda_1=0$ denotes regular motion. In order to compute the MLE we follow the time evolution of a deviation vector $\delta \bm{x}$ from the system's reference orbit $\bm{x}$, and define the finite-time MLE  
\begin{equation}
    \lambda_1 (t) = \frac{1}{t} \ln \frac{\lVert \delta \bm{x}(t) \rVert}{\lVert \delta \bm{x}(0) \rVert},
    \label{eq:finitetimeLEs}
\end{equation}
which quantifies the exponential rate of growth of the deviation vector. In Eq.~\eqref{eq:finitetimeLEs}  $\lVert \cdot \rVert$  stands for the usual Euclidean norm and ${\delta \bm{x}} = \left(\delta x_{1},  \allowbreak \delta x_{2},  \allowbreak \ldots, \delta x_{N},  \allowbreak \delta x_{N+1},\ldots,  \allowbreak \delta x_{2N-1}, \delta x_{2N} \right)$. Furthermore, $t=R \tau$, where $R$ is a positive integer and $\tau$ the renormalization time, i.e., the time interval after which the MLE is computed and the evolved deviation vector is renormalized to its initially considered norm $\lVert \delta \bm{x}(0) \rVert$. Then the MLE $\Lambda_1$ is 
\begin{equation}
    \Lambda_1 = \lim_{t\rightarrow \infty} \lim_{\delta \bm{x} (0) \rightarrow 0} \lambda_1(t).
    \label{eq:LEs}
\end{equation}
As $\Lambda_1$ is unreachable for practical computations, we mainly rely on the evaluation of  $\lambda_1 (t)$ at $t\gg 1$ for its approximation.
Thus, we use the fact that the finite-time MLE evolves as $\lambda_1 (t) \propto t^{-1}\ln t$ for regular trajectories, and approaches a positive constant value for chaotic trajectories.

A numerical technique for the computation of all LEs, based on the time evolution of many deviation vectors $\delta \bm{x}_i$, $i=1, 2, \ldots, k$, with $k \leq 2N$, which are kept linearly independent through a Gram-Schmidt orthonormalization procedure is given in \cite{BGGS1980a,BGGS1980b}. According to this method (which is usually called the \textit{`standard method'}) all other LEs $\Lambda_2$, $\Lambda_3$ etc., apart from the MLE obtained from \eqref{eq:LEs}, are computed as the limits for $t\rightarrow \infty$ and $\delta \bm{x}_i (0) \rightarrow 0$ of some appropriate quantities $\lambda_2(t)$, $\lambda_3(t)$ etc., which are called the finite time LEs (see \cite{BGGS1980b,S2010} for more details).

Another approach for identifying the chaoticity of orbits is obtained by monitoring the alignment between $k$ initially orthogonal deviation vectors. Then the GALI of order $k$ ($\mbox{GALI}_k$) measures the volume of the generalized  parallelepiped whose edges are the corresponding  $k$ unit  vectors $\widehat{\delta \bm{x}}_k(t) = \delta \bm{x}_k (t)/\lVert \delta \bm{x}_k (t)\rVert$, and is given by 
\begin{equation}
    \mbox{GALI}_k(t) = \lVert \widehat{\delta \bm{x}}_1(t) \wedge \widehat{\delta \bm{x}}_2(t) \wedge \ldots \wedge \widehat{\delta \bm{x}}_k(t)\rVert,
    \label{eq:def_GALI_analytical}
\end{equation}
where `$\wedge$' denotes the wedge product of vectors and $t=R\tau$ are the times at which the evolved deviation vectors are again normalized to their initial norm value.

The ability of the GALI method to distinguish between regular and chaotic orbits was theoretically explained in \cite{SBA2007} and has been demonstrated in several applications of the index to various dynamical systems (see e.g.~\cite{HSP2019,Ma2016,Chater2022}).  More specifically, the time evolution of the $\mbox{GALI}_k$ for a regular orbit is~\cite{SBA2007,SM2016}
\begin{equation}
    \mbox{GALI}_k(t) \propto 
    \begin{dcases}
        \phantom{2}\mbox{constant} & \mbox{if $2\leq k \leq N$} \\ 
        \phantom{2}t^{-2(k-N)} & \mbox{if $N < k \leq 2N$}
        \label{eq:behavior_gali_regular_general}
    \end{dcases},
\end{equation}
i.e., the index either remains practically constant for $2\leq k \leq N$, or decays to zero following a power law evolution  for  $N < k \leq 2N$. On the other hand, for chaotic orbits the $\mbox{GALI}_k$ decreases exponentially fast to zero with an exponent which depends on the values of the $k$ largest LEs $\lambda_1\geq \lambda_2 \geq\ \ldots \geq \lambda_k$ \cite{SBA2007}
\begin{equation}
    \mbox{GALI}_k (t) \propto \exp\left(-t\sum_{i=2}^{k}(\lambda_1 - \lambda_i)\right).
    \label{eq:behavior_gali_chaotic_general}
 \end{equation}

\section{Theoretical estimations of the generalized alignment index  numerical evaluation}
\label{sec:theoretical_estimate}

Our aim is to investigate the numerical errors arising during the numerical computation of the GALI using the MPM and the VM. Before proceeding we outline a number of assumptions aiming to simplify our analysis. In the rest of this section we assume that the volume formed by the unit tangent vectors to the reference orbit, i.e.,~the $\mbox{GALI}$ value at any time of the system's evolution, is bounded by the volume  of a right angled parallelepiped with sides the same unit tangent vectors. It follows that it is enough to focus on the product of the magnitudes of the tangent vectors as it shares the same numerical error with the $\mbox{GALI}$ up to a multiplicative factor which depends on the orientations between the different tangent vectors.

Furthermore, given that all the tangent vectors to the reference orbit have initially the same magnitude, and due to the fact that they are evolved independently, we  consider that the numerical error accumulates equally for all  tangent vectors. That is to say, it is sufficient to evaluate the numerical error of one single vector to deduce the numerical error during the computation of the volume formed by all  tangent vectors, which is proportional to (or of the same order as) the $\mbox{GALI}$'s error. 

In addition, we roughly consider that the origin of the numerical errors stems from three main contributions which are independent for all times: (i) the machine precision, (ii) the magnitude of the tangent vectors at the renormalization time, and  (iii)  the global truncation errors of the numerical integration schemes~\cite{MH2018}.
Therefore, the numerical error in the computation of the $\mbox{GALI}_k$ is intricate and can be represented as a multivariate polynomial of order $k$ with constant coefficients, whose variables correspond to the control parameters above.
In order to analyze the error meaningfully, we consequently focus only on the leading order terms, because they are simple and depend on the main {\it extensive parameters} influencing the numerical evolution of the phase space orbits and their associated tangent vectors.
We will show that such terms are the primary contributors to the numerical inaccuracies of the computed $\mbox{GALI}$ in short-time scales.

For completeness, we point toward additional notes relevant to this work in the literature. The effects of the global truncation errors of the numerical integration scheme, the length of the initial tangent vectors, the renormalization time and the machine precision on the computation of the MLE are investigated in Ref.~\cite{MH2018}. Furthermore, in Sect.~3.5 of Ref.~\cite{PP2016} (and references therein), additional sources of numerical errors in the computation of the LEs are studied. 
For instance, the numerical orthogonalization's related errors and the statistical errors due to the finite time computations of the LEs. In the case of the GALI method, the effects of the numerical integration techniques during its numerical computation are briefly investigated in Ref.~\cite{GES2012}.

\subsection{The variational method}
\label{subsec:variational_method}

In this section, we focus on evaluating the numerical uncertainty of the computation of the GALI using the variational approach.
It is worth noting that the VM constitutes the sole approach considered in previous studies (see e.g.~\cite{SBA2007,SM2016}) for the computation of the GALI. Thus, later on, we will prove that the definition of the GALI in the variational framework is equivalent to the definition using the MPM. The variational equations are obtained by linearizing the equations of motion [Eq.~\eqref{eq:equation_of_motion_general}] around a reference orbit $\bm{x}(t)$ in the phase space.
Considering a set of $k$ orthogonal deviation (tangent) vectors 
\begin{align*}
    \bm{w}_i(t_0) &= {\delta \bm{x}}_i (t_0) = \Big( \delta x_{i, 1}(t_0), \delta x_{i, 2}(t_0),\ldots, \delta x_{i, N}(t_0),\\
    &\qquad \delta x_{i, N+1}(t_0),\ldots, \delta x_{i, 2N-1}(t_0), \delta x_{i, 2N}(t_0) \Big)\\
    &= \Big( w_{i, 1}(t_0), w_{i, 2}(t_0), \ldots,  w_{i, N}(t_0),\\
    &\qquad w_{i, N + 1}(t_0), w_{i, N+2}(t_0), \ldots, w_{i, 2N}(t_0) \Big)
\end{align*}
to the system's orbit $\bm{x}_0$ in phase space, the variational equations are made up of $2kN$ differential forms
\begin{equation}
  \dot{\bm{w}}_i = \bm{J}_{2N}\cdot \bm{D} ^2 \mathcal{H}\left(\bm{x}(t)\right) \cdot \bm{w}_i = \frac{\partial \bm{f} \left(\bm{x}(t)\right)}{\partial \bm{x}} \cdot \bm{w}_i = \bm{A}(t) \cdot \bm{w}_i,
  \label{eq:variational_equation_general}
\end{equation}
where $\bm{D}^2\mathcal{H}\left(\bm{x}(t)\right)$ is the Hessian matrix of the Hamiltonian evaluated at the reference orbit $\bm{x}(t)$, and the subscript $i$ indexes the tangent vector. As initial conditions for the variational equations, the common practice is to choose random coordinates for the deviation vectors, which are then rescaled in such a way that they all share the same norm, i.e., $\lVert \bm{w}_i(0)\rVert =\lVert \bm{w}_{i,0} \rVert = d_0$ ($d_0>0$).

The formal solution of Eq.~\eqref{eq:variational_equation_general} reads
\begin{equation}
  \bm{w}_i(t) = \exp \left[\int_{0}^{t} \bm{A}(t^\prime)dt^\prime\right] \cdot \bm{w}_i(0).
  \label{eq:formal_solution_variational_equation}
\end{equation}
In what follows, we first assume that the trajectory of the system $\bm{x}(t)$ and its tangent dynamics $\bm{w}_i(t)$ are known in a closed form. Consequently, we define the GALI of order $k$ (denoted as $G_k$ for the VM) of the orbit $\bm{x}$ after a renormalization time $\tau$~\cite{SBA2007,SM2016} as
\begin{equation}
  \mbox{G}_k (\tau) = \lVert \bm{u}_1 (\tau) \wedge \bm{u}_2 (\tau) \wedge \ldots \wedge \bm{u}_k(\tau) \rVert, 
  \label{eq:gali_variational_equation}
\end{equation}  
where   
\begin{equation}
   \bm{u}_i(\tau) = \frac{\bm{w}_i(\tau)}{\lVert \bm{w}_i (\tau)\rVert},
    \label{eq:renormalization_dvd}
\end{equation}
is the unit vector of the $i^\textit{th}$ deviation vector $\bm{w}_i(\tau)$.

We first investigate the error due to the machine precision (referred to as $\varepsilon$) during the computation of the $\mbox{GALI}_k$ after one renormalization time $\tau$, which we denote as $G_k (\tau, \varepsilon)$. Starting from unit deviation vectors, i.e.,~$d_0 = 1$, and assuming that the renormalization time $\tau$ is small, we have $\lVert \bm{w}_i(\tau)\rVert \approx 1$ such that 
\begin{equation}
    \widetilde{u}_{i,m} \left(\tau \right)= u_{i, m}\left(\tau \right) + \mathcal{O}\left(\varepsilon\right),
    \label{eq:machine_prec_vm_01}
\end{equation}
with $\widetilde{u}_{i,m}$ denoting the machine representation of $u_{i,m}$. In general, we will denote by $\widetilde{Q}$ any approximation of the variable $Q$. Thus, the computer representation of the GALI of order $k$ is 
\begin{equation}
  \widetilde{G} _k (\tau, \varepsilon) = \lVert \widetilde{\bm{u}} _1 (\tau) \wedge \widetilde{\bm{u}}_2 (\tau) \wedge \ldots \wedge \widetilde{\bm{u}}_k(\tau) \rVert, 
  \label{eq:gali_variational_equation_02}
\end{equation}  
in such as way that the error due to the machine precision becomes
\begin{equation}
  \widetilde{G}_k (\tau, \varepsilon) - G_k(\tau) = \mathcal{O}\left(\varepsilon^k\right).\label{eq:gali_renormalization_variational_equation}
\end{equation}

In addition to the effect of the machine precision on the computation of the $\mbox{GALI}$, the magnitude of the initial deviation vectors, $d_0 = \lVert \bm{w}_i (t_0)\rVert$, also impacts the computed GALI values at each renormalization time. By denoting with $\widetilde{G}_k \left(\tau, \rho\right)$ the GALI$_k$  obtained by replacing $\bm{w}_i (t_0)$ by $\bm{w}_i^{\rho} (t_0) = \rho \cdot \lVert \bm{w}_i (t_0)\rVert$, where $\rho = d_0/\lVert \bm{w}_i (t_0)\rVert$ is the rescaling factor, we see that the time evolution of the rescaled deviation vectors $\bm{w}_i^{\rho}(t)$ is given by 
\begin{equation}
  \bm{w}_i^{\rho} (t) = \rho \bm{w}_i(t),
  \label{eq:rescaled_deviation_variational_equation}
\end{equation} 
with the time evolution of $\bm{w}_i(t)$ being generated by the solution to the variational equations [Eq.~\eqref{eq:formal_solution_variational_equation}].
Consequently,
\begin{equation}
  \bm{u}_i^{\rho}(\tau) = \frac{\bm{w}_i^{\rho}(\tau)}{\lVert \bm{w}_i^{\rho}(\tau) \rVert} = \frac{\rho \bm{w}_i(\tau)}{\lVert \rho \bm{w}_i(\tau) \rVert} = \frac{\bm{w}_i(\tau)}{\lVert \bm{w}_i(\tau) \rVert} = \bm{u}_i(\tau),
  \label{eq:rescaled_deviation_variational_equation_02}
\end{equation}
and the magnitude of the $k$ initial deviation vectors do not affect the calculations of GALI in case of the VM. In this context, we only have to take into account round off errors for the computer representation of $\bm{u}_i^{\rho}$, similar to what was done in Eq.~\eqref{eq:machine_prec_vm_01}. Thus, the numerical inaccuracy of the GALI$_k$ at the renormalization time due to the magnitude of the deviation vectors leads to
\begin{equation}
  \widetilde{G}_k(\tau, \rho) - G_k(\tau) = \mathcal{O}\left(\varepsilon^k\right).
  \label{eq:gali_rescaled_variational_equation}
\end{equation}

Combining the errors due to machine precision [Eq.~\eqref{eq:gali_renormalization_variational_equation}] and the ones due to the magnitude of the deviation vectors at the renormalization time [Eq.~\eqref{eq:gali_rescaled_variational_equation}] leads to
\begin{equation}
  \widetilde{G}_k(\tau) - G_k(\tau)  = \mathcal{O}\left(\varepsilon^k\right).
  \label{eq:gali_variational_equation_00}
\end{equation}

\subsection{The multi-particle method}
\label{subsec:twoparticle_method}

We can also define the $\mbox{GALI}_k$ in the framework of the MPM.
Starting from the initial condition $\bm{x}(t_0) = \bm{x}_0$ and the $k$ initially orthogonal separation (tangent) vectors 
\begin{align*}
    \bm{d}_i(t_0) &= {\delta \bm{x}}_i  = (\delta x_{i, 1}(t_0), \delta x_{i, 2}(t_0),\ldots, \delta x_{i, N}(t_0),\\
    &\qquad \delta x_{i, N+1}(t_0),\ldots, \delta x_{i, 2N-1}(t_0), \delta x_{i, 2N}(t_0))\\
    &= (d_{i, 1}(t_0), d_{i, 2}(t_0), \ldots,  d_{i, N}(t_0),\\
    &\qquad d_{i, N + 1}(t_0), d_{i, N+2}(t_0), \ldots, d_{i, 2N}(t_0))
\end{align*}
with norm $\lVert \bm{d}_i(t_0)\rVert = d_0$, the orbit $\bm{x}(t)$ and its neighboring orbits $\bm{y}_i(t) = \bm{x}(t) + \bm{d} _i (t)$ can be obtained by integrating the equations of motion [Eq.~\eqref{eq:equation_of_motion_general}] $k + 1$ times. In this context, the GALI  of order $k$ (referred to as $J_k$ for the MPM) is defined as
\begin{equation}
  J_k(\tau) = \lVert \bm{v}_1(\tau) \wedge \bm{v}_2(\tau) \wedge \ldots \wedge \bm{v}_k(\tau) \rVert, 
  \label{eq:gali_two_particle}
\end{equation}
at every renormalization time $\tau$, with 
\begin{equation}
  \bm{v}_i (\tau) = \frac{\bm{d} _i (\tau)}{\lVert \bm{d} _i (\tau) \rVert},
  \label{eq:renormalization_two_particle}
\end{equation}
the unit vector of the $i^\textit{th}$ separation vector in the phase space. Our first task here is to show that Eq.~\eqref{eq:gali_two_particle} is equivalent to Eq.~\eqref{eq:gali_variational_equation} i.e., $J_k (\tau) = G_k(\tau)$, so that the GALIs defined using the VM and the MPM are equal.

In order to achieve our goal, we must find ways to approximate the evolution of the $\bm{d}_i(0)$. We use a Taylor series expansion  
\begin{equation}
  \dot{\bm{d}}_i \approx \bm{A}(t) \bm{d} _i + \frac{\bm{D}^{2} \bm{f}(\bm{x})}{2} \bm{d}_i^2
  + \mathcal{O}\left(\lVert\bm{d}\rVert^3\right),
\label{eq:two_particle_equation_genaral_02}
\end{equation} 
to approximate the time evolution of $\bm{d}_i$, with $\mathcal{O}\left(\lVert \bm{d}_i \rVert^3\right) \rightarrow 0$, because  $\lVert \bm{d}_i \rVert \ll 1$. At this point, it is convenient for us to approximate the evolution of each element $d_{i, m}$ of the separation vector with index $i$, by the dynamics of its fastest changing coordinate corresponding to the direction of maximal stretching of the tangent vectors, whose rate of growth is the MLE~\cite{S2010}. Thus Eq.~\eqref{eq:two_particle_equation_genaral_02} can be simplified to
\begin{equation}
  \dot{d}_{i,m}(t) \approx \lambda_1 d_{i,m}(t) + \Gamma d_{i,m}^2(t),
  \label{eq:two_particle_equation_general_03}
\end{equation}
with $\lambda_1$ being the MLE and $\Gamma$ a scaling coefficient of the second order derivative in the Taylor expansion [Eq.~\eqref{eq:two_particle_equation_genaral_02}]. Without loss of generality, we fix $\Gamma = \mathcal{O}(1)$.

Equation~\eqref{eq:two_particle_equation_general_03} can be solved analytically, considering an initial perturbation $d_{i,m} (t_0) \propto d_0$ at time $t_0 =0$. Its solution can be written as
\begin{equation}
  d_{i,m}(t) \approx \frac{\lambda_1 d _0 e^{\lambda_1 t}}{\lambda_1  + d _0 \left(1 - e^{\lambda_1 t}\right)},
\end{equation}
which can be simplified to
\begin{equation}
  d_{i,m}(t) \approx d _0 e^{\lambda_1 t} + d _0^2 t e^{\lambda_1 t} + \frac{1}{2}d _0^2 \lambda_1 t^2e^{\lambda_1 t} + \mathcal{O}\left(d_0^3 t^2 e^{\lambda_1t} \right),
  \label{eq:separation_evolution_02}
\end{equation}
with $\mathcal{O}\left(d_0^3 t^2 e^{\lambda_1t} \right)\rightarrow 0$ again because $d_0 \ll 1$. It follows that, the first term on the right hand side of Eq.~\eqref{eq:separation_evolution_02} represents the asymptotic solution [Eq.~\eqref{eq:formal_solution_variational_equation}] of the variational equations [Eq.~\eqref{eq:variational_equation_general}] in the limit of large time, as also pointed out for example in Eq.~(30) of Ref.~\cite{SBA2007}, i.e., $\bm{w}_i (t) \approx \bm{w}_0 e^{\lambda_1 t}$. As a result, we write
\begin{equation}
  d_{i, m} (t) \approx w_{i, m} (t) + d _0^2t e^{\lambda_1 t} + \frac{1}{2}d _0^2  \lambda_1 t^2e^{\lambda_1 t}.
  \label{eq:separation_evolution_03}
\end{equation}

Now let us turn back to expressing $J_k$~[Eq.~\eqref{eq:gali_two_particle}] after one renormalization time $\tau$. In order to perform that task, we note that for small enough values of $\tau$, we have  $\lVert \bm{\delta} _i (\tau)\rVert \approx  \lVert \bm{w} _i (\tau)\rVert \approx d_0$ leading to 
\begin{equation}
  v_{i, m} (\tau) \approx \frac{d_{i, m}(\tau)}{d_0}, \quad u_{i, m} (\tau) \approx \frac{w_{i, m} (\tau)}{d_0}.
  \label{eq:two_particle_renormalization_02}
\end{equation}
Consequently, based on Eq.~\eqref{eq:separation_evolution_03} and Eq.~\eqref{eq:two_particle_renormalization_02}, we write
\begin{equation}
  \bm{v}_i (\tau) - \bm{w}_i(\tau) \approx \frac{1}{d_0} \bm{\Omega},
\end{equation}
where $\bm{\Omega}$ is the vector with coordinates $\Omega_m \approx d _0^2 \tau e^{\lambda_1 \tau} + \frac{1}{2}d _0^2 \lambda_1 \tau^2e^{\lambda_1 \tau}$. This results in
\begin{equation}
  J_k (\tau) \approx G_k (\tau) + \mathcal{O}\left(d_0^k \tau^k \right) + \mathcal{O}\left(d_0^k \tau^{2k}\right),
  \label{eq:gali_two_particle_01}
\end{equation}
where we consider only leading order terms. Therefore, it turns out that for $d_0 \rightarrow 0$, the GALI$_k$ as defined using the VM and the MPM practically coincide for fixed $\tau$.

Let us now look at the different numerical inaccuracies in computing the GALI$_k$ in the case of the MPM [Eq.~\eqref{eq:gali_two_particle}]. We first investigate the influence of machine precision when computing the GALI$_k$ after one renormalization time $\tau$, i.e.,
\begin{equation}
  \widetilde{J}_k(\tau, \varepsilon) = \lVert \widetilde{\bm{v}}_1(\tau) \wedge \widetilde{\bm{v}}_2(\tau) \wedge \ldots \wedge \widetilde{\bm{v}}_k(\tau) \rVert .
  \label{eq:gali_two_particle_02}
\end{equation}
Clearly, as for the VM [see Eq.~\eqref{eq:gali_variational_equation_02}], we find
\begin{equation}
  \widetilde{J}_k (\tau, \varepsilon) - J_k(\tau) = \mathcal{O}\left(\varepsilon^k\right).
\end{equation}

The magnitude of the initial separation vectors at the renormalization time [Eq.~\eqref{eq:renormalization_two_particle}] also impacts the computation of the GALI by  the MPM. If we assume that the reference orbit $\bm{x}(t)$ is bounded (i.e.,~an orbit which does not escape to infinity) then $\widetilde{\bm{x}}_m(t) = \bm{x}_m(t) + \mathcal{O}(\varepsilon)$ for all time, which also applies for all neighboring orbits $\bm{y}_i(t)$. As a result, 
\begin{equation}
  \widetilde{d}_{i,,m}(t) = \widetilde{y}_{i,m}(t) - \widetilde{x}_m(t) = d_{i, m}(t) + \mathcal{O}(\varepsilon).
  \label{eq:two_particle_separation_machine_precision}
\end{equation}
with $\bm{y}_{i,m}$ being the phase space coordinates of the $i^{th}$ nearby orbit from the reference orbit with coordinate $x_m$. Consequently, the renormalization procedure [Eq.~\eqref{eq:renormalization_two_particle}] gives
\begin{equation}
  \widetilde{\bm{v}}_i(\tau) = \frac{\widetilde{\bm{d}_i}(\tau)}{\lVert \widetilde{\bm{d}_i}(\tau)\rVert} \approx
  \frac{\bm{d}_i (\tau) + \mathcal{O}(\varepsilon)}{d_0} = \bm{v}_i (\tau) + \mathcal{O}\left(\frac{\varepsilon}{d_0}\right),
\end{equation}
so that the leading order term in the numerical error of the GALI$_k$ due to the magnitude of the separation vector becomes
\begin{equation}
  \widetilde{J}_k(\tau, \rho) - J_k(\tau) = \mathcal{O}\left(\frac{\varepsilon^k}{d_0^k}\right).
  \label{eq:gali_two_particle_03}
\end{equation}

Gathering together Eq.~\eqref{eq:gali_two_particle_01}, Eq.~\eqref{eq:gali_two_particle_02} and Eq.~\eqref{eq:gali_two_particle_03} leads to 
\begin{equation}
  \widetilde{J}_k(\tau) - G_k(\tau) = \mathcal{O}\left(d_0^k \tau^k \right) + \mathcal{O}\left(d_0^k \tau^{2k}\right) + \mathcal{O}\left(\frac{\varepsilon^k}{d_0^k}\right) + \mathcal{O}\left(\varepsilon^k\right).
  \label{eq:gali_two_particle_equation_00}
\end{equation}

\subsection{Dependence of the generalized alignment index values on the global numerical truncation errors}
\label{subsec:theo_numerical_truncation}

In the previous calculations, we worked assuming that we know the analytical form of the solution of the phase space and its associated tangent dynamics [Eq.~\eqref{eq:equation_of_motion_general}]. Nevertheless, for most models in nonlinear dynamics, the exact solution cannot be obtained. Therefore, we approximate the analytical solution of the system by a numerical one so that truncation errors introduced by the numerical integration scheme during the evolution of the orbit and its tangent space can be accounted for in the computation of the GALI.  

In Ref.~\cite{MH2018}, it was argued that a numerical integration scheme of order $p$, with fixed integration time step $h$, affects the time propagation of the tangent vectors at time $T$ by a term $\varepsilon_T = \mathcal{O}\left( h^p\right)$. This term accounts for the global truncation error on each tangent vector at time $T$. Then, the numerical error of the GALI$_k$ when we include the global truncation errors can be obtained following the same steps as above, leading to
\begin{equation}
  \widetilde{G}_{k} - \mbox{GALI}_k = \mathcal{O}(\varepsilon^k) + \mathcal{O}\left(\varepsilon_T^k\right),
  \label{eq:gali_variational_method_final}
\end{equation}
when computing GALI using the VM. On the other hand, when the MPM is used, we have
\begin{equation}
  \widetilde{J}_{k} - \mbox{GALI}_k = \mathcal{O}\left(d_0^k \tau^k\right) + \mathcal{O}\left(d_0^k\tau^{2k}\right) + \mathcal{O}\left(\frac{\varepsilon^k}{d_0^k}\right) + \mathcal{O}(\varepsilon^k) + \mathcal{O}\left(\varepsilon_T^k\right).
  \label{eq:gali_two_particle_method_final}
\end{equation}

\begin{figure*}
    \centering
    \includegraphics[width=0.85\textwidth]{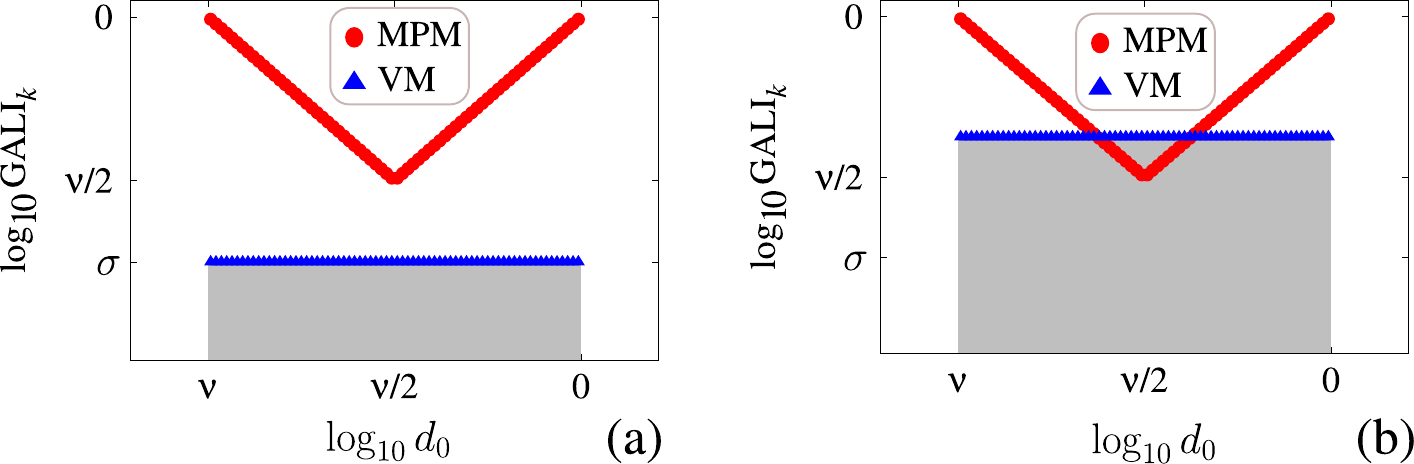}
    \caption{Theoretical reliable regions for the computation of the GALI of order $k$ by  the  VM (blue triangles) and the  MPM (red circles) for unit renormalization time $\tau = 1$.     The labels are as follows: $\nu=log_{10}\varepsilon$, $d_0=\lVert w_0 \rVert = \lVert \delta _0 \rVert $, $\sigma = \log_{10} \varepsilon_T$.   Panel (a): $\varepsilon \lesssim \varepsilon_T \lesssim \varepsilon^{1/2}$.    Panel (b): $\varepsilon^{1/2}\lesssim \varepsilon_T $.
    }
    \label{fig:GALI_ft_do_theoretical predictions}
\end{figure*}

In Fig.~\ref{fig:GALI_ft_do_theoretical predictions} we plot the dependence of the $\mbox{GALI}_k$ against $d_0$ at fixed $\tau=1$ value. The diagrams obtained define the reliable regions of $\mbox{GALI}_k$, by estimating the expressions of Eq.~\eqref{eq:gali_variational_method_final} (blue filled-triangles) and Eq.~\eqref{eq:gali_two_particle_method_final} (red filled-circles). More specifically, we examine two cases namely when $\varepsilon < \varepsilon_T < \varepsilon^{1/2}$ shown in Fig.~\ref{fig:GALI_ft_do_theoretical predictions}(a), and $\varepsilon_T > \varepsilon^{1/2}$ in Fig.~\ref{fig:GALI_ft_do_theoretical predictions}(b). Note that in Fig.~\ref{fig:GALI_ft_do_theoretical predictions} $\nu=\log_{10}\varepsilon$ and $\sigma=\log_{10}\varepsilon_T$. Overall, we see that reliable regions for the computation of the GALI are independent of the GALI order. In addition the $\mbox{GALI}_k$ computed using the VM is independent of the magnitude of the initial tangent vectors $d_0$. In fact, in this case the computed values of the GALI only depend on the global truncation error $\varepsilon_T$ of the numerical integration scheme, since the latter is, in general, much larger than the machine precision $\varepsilon$. On the other hand, the results of the reliable regions using the MPM can be separated in two cases. Whenever $\varepsilon_T \in [\varepsilon, \varepsilon^{1/2}]$, we obtain a $V-$shape with minimum located at $d_0 \sim \varepsilon^{1/2}$ [red-colored shape in Fig.~\ref{fig:GALI_ft_do_theoretical predictions}(a)]. Furthermore, when $\varepsilon_T > \varepsilon$, the $V-$shape loses its pointy edge to a flat bottom, which means that the most accurate results using the MPM are also dependent on the $\varepsilon_T$ value [Fig.~\ref{fig:GALI_ft_do_theoretical predictions}(b)]. 

\section{Numerical results}
\label{sec:numerical_results}

\subsection{Hamiltonian models  and computational setup}
\label{sec:model_descriptions}

The first model we study is the well known,  two degree of freedom (2D) HH system, whose Hamiltonian function reads~\cite{HH1964}
\begin{equation}
    \mathcal{H} = \frac{1}{2}\left(p_1^2 + p_2^2\right) + \frac{1}{2} \left(x_1^2 + x_2^2\right) + x_1^2x_2-\frac{1}{3}x_2^3,
    \label{eq:hamiltonian_hh}
\end{equation}
with $x_1$ and $x_2$ being the system's generalized positions, and $p_1$ and $p_2$ their associated canonical momenta. This model gives a generic description of interactions between two asymmetric celestial bodies whose motion is restricted to a two dimensional plane (e.g., the motion of a planet around its sun, a star around the center of its galaxy, etc.) or molecules with asymmetric distributions of charges. As an additional remark we note that the value of the Hamiltonian function itself tunes the strength of the system's nonlinearity.

The second model we investigate is the multidimensional $\beta$-FPUT model which consists of a chain of $N$ coupled nonlinear oscillators.
Its Hamiltonian function is expressed as~\cite{FORD1992}
\begin{equation}
    \label{eq:bFPUT}
    \mathcal{H}_N = \frac{1}{2} \sum_{i=1}^{N}p_i^2 + \sum_{i=0}^{N}\left[\frac{1}{2}\left(x_{i+1}-x_{i}\right)^2 + \frac{\beta}{4}\left(x_{i+1}-x_{i}\right)^4\right],
\end{equation}
where $x_i$ is the displacement of the $i^\textit{th}$ oscillator, $p_i$ its conjugate momentum, and the parameter $\beta$ controls the strength of the nonlinear terms. The $\beta$-FPUT model is used in preference to the $\alpha$-FPUT due to the absence of escaping orbits in the former, which greatly simplifies our work. The FPUT system is a generic central model in the wide and intensive field of nonlinear lattice dynamics. As such, it has been considered in investigations of heat conduction~\cite{PMDN2004}, thermalization~\cite{OVPL2015}, ergodicity~\cite{DCF2017}, solitons~\cite{ZWZZ2005}, breathers~\cite{FG2008}, classical and quantum chaos~\cite{CDRT1998,YR2024} and the possible connection between the aforementioned physical processes, see for example~\cite{BI2005} and references therein. In our numerical simulations, we set $\beta=1$ and tune the nonlinearity of the system by changing the value $\mathcal{H}_N$ of the energy. Furthermore, fixed boundary conditions (i.e., $x_0=x_{N+1}=0$) are imposed at the two edges of the lattice.

\begin{figure}[h!]
    \centering
    \includegraphics[width=\columnwidth]{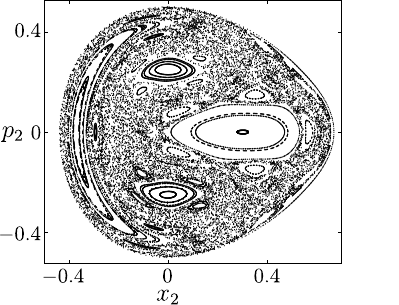}
    \caption{
    {\bf 2D HH system.} 
    The PSS of the HH model [Eq.~\eqref{eq:hamiltonian_hh}] at $x_1 = 0$ for $\mathcal{H} = 0.125$.
    For each point $(x_2,p_2)$ of the PSS, the remaining coordinate $p_1$ is found such that $p_1>0$.
    }
    \label{fig:pss_hh_h=0.125}
\end{figure}

These two models  provide a perfect setup to investigate the behavior of chaos indicators for different dynamical regimes encountered in nonlinear systems. In that regard they are extremely appreciated in computational physics, see e.g.~\cite{SG2010,GES2012,MH2018}. In the case of the HH, an initial condition of a specific nature can be identified once we generate the system's Poincar\'e surface of section (PSS) associated with a particular $\mathcal{H}$ [Eq.~\eqref{eq:hamiltonian_hh}]. Figure~\ref{fig:pss_hh_h=0.125} depicts an example of a PSS defined by $x_1=0$ and $p_1>0$ with $\mathcal{H}=0.125$.

Turning now to the $\beta$-FPUT lattice model, the determination of the orbits with a specific dynamical nature is more complex mainly due to the high dimensionality of the system's phase space. Fortunately, there exist orbits which are rather simple to locate and whose vicinity has been extensively studied, the so-called {\it simple periodic orbits} of order 1 (SPO1)~\cite{ABS06,AB2006}. Thus in our study, we work with perturbations of these SPO1 orbits.
In short, the SPO1 is a nonlinear continuation of the mode of the linearized system with wave number $\nu=(N+1)/2$, taking the form
\begin{equation}
    x_{2j}(0) = 0, \quad x_{2j-1}(0) = -x_{2j+1}(0), 
    \label{eq:bfput_spo1}
\end{equation}
for $j = 1, 2, \ldots, \frac{N+1}{2}$ with the momentum of all particles set to $p_j(0)=0$. Given a finite number of oscillators the SPO1 orbits are regular for energy values $\mathcal{H}_N$ below the so-called first destabilization energy threshold $\mathcal{H}_N^{c}$, while being chaotic past that threshold~\cite{AB2006}. The $\mathcal{H}_N^{c}$ can be found using the standard linear stability analysis (see~\cite{S2001,D2004,FG2008,MMS2020} for details). 

Let us now explain how we practically compute the LEs and the GALIs.
To estimate the first $k$ largest LEs, we integrate simultaneously the equations of motion along with a set of $k$ variational equations or nearby-orbits. For each simulation, we start from $k$ random and linearly independent tangent vectors [$\bm{d}_i(0)$ or $\bm{w}_i(0)$] which are rescaled to achieve the size $d_0$.  Then, the LEs are estimated using the `standard method'~\cite{BGGS1980a,BGGS1980b,S2010}, leaving the renormalization time $\tau$, and the deviation vector size $d_0$, as free parameters.  A pseudo-code for the implementation of VM for this approach can be found in~\cite{S2010}. For the sake of completeness, in~\ref{app:sec:pseudo_code} we also provide a pseudo-code for the computation of the GALI using the MPM, which is based on the approach developed in~\cite{SBA2008}  (see Ref.~\cite{SM2016} for the VM counterpart).

We evolve the system's phase space orbits and the associated neighboring trajectories using the symplectic  integration scheme $ABA864$ of order $4$~\cite{BCFLMM2013,FLBCMM2013,SS2018,DMMS2019} with fixed time step, $h$.
The value of $h$ controls the global truncation error $\varepsilon_T$ which we track at every time, $t>0$, using the {\it relative energy error} 
\begin{equation}
    E_r(t)= \left\lvert \frac{\mathcal{H}(t) - \mathcal{H}(0)}{\mathcal{H}(0)} \right\rvert, \quad
    E_r^N(t)= \left\lvert \frac{\mathcal{H}_N(t) - \mathcal{H}_N(0)}{\mathcal{H}_N(0)} \right\rvert, 
    \label{eq:relative_energy_error}
\end{equation}
for the HH and the $\beta$-FPUT models respectively.
In the case of the VM the tested orbit and the related variational equations are simultaneously integrated by the $ABA864$ integrator using the so-called `tangent map method'~\cite{SG2010,GES2012}.
In addition, we perform double precision computations, i.e.,~$\varepsilon \approx 10^{-16}$, as this is the default numeric type of almost all modern computational tasks.

\subsection{The H\'enon-Heiles model}
\label{subsec:henonheiles_num}

\begin{figure}[!htbp]
    \centering 
        \includegraphics[width=\columnwidth]{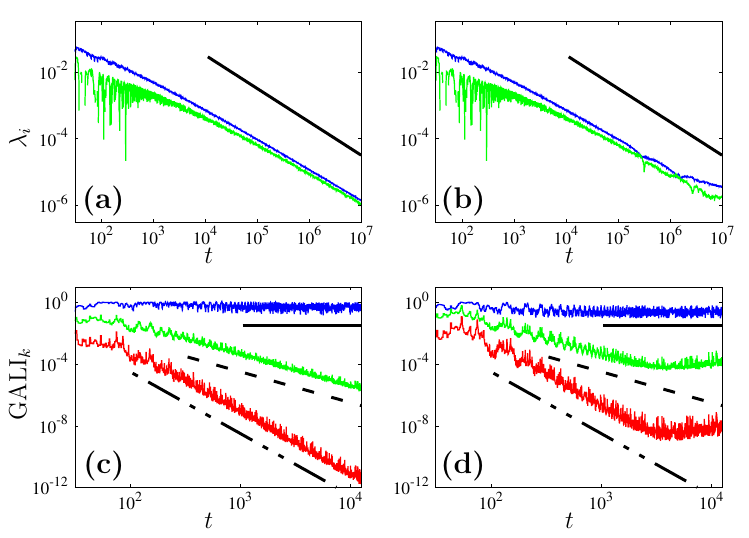}
    \caption{{\bf 2D HH system}. Time evolution of (a-b) the two largest finite-time LEs $\lambda_1$ (blue curves) and $\lambda_2$ (green curves), and (c-d) the $\mbox{GALI}_2$ (blue curves), $\mbox{GALI}_3$ (green curves) and $\mbox{GALI}_4$ (red curves) for a regular orbit with initial conditions  $x_1=0$, $p_1>0$, $x_2=0.558$ and $p_2=0$ of the HH with $\mathcal{H} = 0.125$. The results are obtained by (a), (c) the VM and (b), (d) the MPM. The continuous black lines in (a-b) guide the eye for slope $-1$, while the continuous, dashed, and dashed-dotted ones in (c-d) have slopes $0$ (horizontal lines), $-2$ and $-4$ respectively. All the calculations are performed for $d_0=10^{-8}$.
     }
    \label{fig:R1_MLE_GALI_Rep_01}
\end{figure}

We start our study by demonstrating that the computation of the GALIs  using the VM and the MPM genuinely give the same results by considering specific initial conditions in Fig.~\ref{fig:pss_hh_h=0.125}. In Figs.~\ref{fig:R1_MLE_GALI_Rep_01} (a-b), we plot the time dependence of the two largest LEs $\lambda_1$ (blue curves) and $\lambda_2$ (green curves),  for the regular orbit with initial conditions $x_1=0$, $x_2=0.558$, $p_2=0$ and $p_1>0$ evaluated such that $\mathcal{H}=0.125$. Since we also require initial conditions for the tangent vectors, randomly selected coordinates drawn from a uniform distribution are used, and the created deviation vectors are orthogonalized and rescaled to fit the desired magnitude for all tangent vectors $d_0=10^{-8}$. Note that we scale the deviation vectors in the VM computations as good practice for comparison with the MPM.

The finite-time LEs are computed using both the VM [Fig.~\ref{fig:R1_MLE_GALI_Rep_01}(a)] and the MPM [Fig.~\ref{fig:R1_MLE_GALI_Rep_01}(b)]. In our simulations we integrate equations associated with each method up to a final time $t=10^7$,  setting the integration time step to $h=0.01$, which ensures that the relative energy error [Eq.~\eqref{eq:relative_energy_error}] is upper bounded by $E_r \approx 10^{-11}$. Furthermore, we fix the renormalization time to $\tau = 1$. For both methods, the computed LEs are decaying following the power law $\lambda_{1,2} \propto  t^{-1}$ [black lines in Figs.~\ref{fig:R1_MLE_GALI_Rep_01}(a-b)]. This behavior confirms that the studied orbit is regular. In addition, a divergence from the $t^{-1}$ power law of the computed LEs through of the MPM is observed after $t \approx 10^{6}$ for which the magnitudes of the LEs are $\lambda_{1,2}\approx 10^{-6}$. This divergence suggests a limit in the precision of the computation of the LEs using the MPM for regular orbits~\cite{MH2018}.

Using the same model parameters, we also compute the time evolution of the GALIs. The numerical results of the time evolution of $\mbox{GALI}_2$ (blue curve), $\mbox{GALI}_3$ (green curve) and $\mbox{GALI}_4$ (red curve) are shown in Fig.~\ref{fig:R1_MLE_GALI_Rep_01}(c) for the VM.  In particular, we see that the VM is quite accurate in capturing the theoretical decay rates with slopes $-2$ and $-4$, for the GALI$_3$ and the GALI$_4$ respectively, which are indicated by  the dashed and dashed-dotted black lines in Fig.~\ref{fig:R1_MLE_GALI_Rep_01}(c), as well as the practical constancy of GALI$_2$ [horizontal black line in Fig.~\ref{fig:R1_MLE_GALI_Rep_01}(c)].  The novelty of our work, is that we perform the same computations of the GALIs, using the MPM and demonstrate the effectiveness of this approach.

In Fig.~\ref{fig:R1_MLE_GALI_Rep_01}(d) we show the time evolution of the $\mbox{GALI}_2$ (blue curve), $\mbox{GALI}_3$ (green curve) and $\mbox{GALI}_4$ (red curve) computed using the MPM with the same set of tangent vectors used in the VM computations of Fig.~\ref{fig:R1_MLE_GALI_Rep_01}(c). There we see  a practically constant $\mbox{GALI}_2$ which remains similar to the one obtained using the VM [blue curve of Fig.~\ref{fig:R1_MLE_GALI_Rep_01}(c)]. On the other hand, for both the $\mbox{GALI}_3$ and the $\mbox{GALI}_4$ an interesting behavior is seen. At the early stage of the evolution, up to $t\lesssim 10^{3.5}$, the computed $\mbox{GALI}_3$ and $\mbox{GALI}_4$ are decaying following the same power laws [respectively denoted by the dashed and dashed-dotted black lines in Fig.~\ref{fig:R1_MLE_GALI_Rep_01}(d)] as for the VM. Then, for $t \gtrsim 10^{3.5}$, a leveling off to a practically constant value appears, taking the values of the computed $\mbox{GALIs}$ away from their expected power laws. As in the case of the LEs computations, that time marks the point from which the accumulations of numerical errors dominates the numerical values of the computed GALIs.

\begin{figure}[h!]
    \centering 
    \includegraphics[width=\columnwidth]{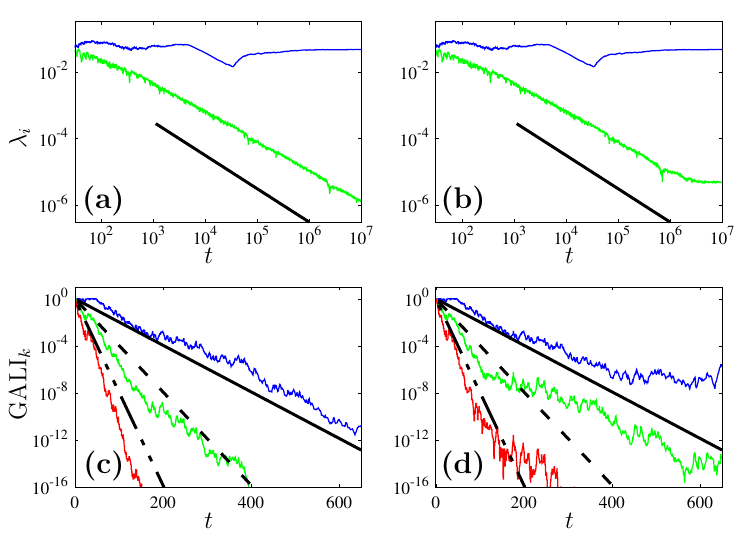}
    \caption{{\bf 2D HH system}. Similar to Fig.~\ref{fig:R1_MLE_GALI_Rep_01}, but for a chaotic orbit with initial coordinates $x_1=0$, $p_1=0.42$, $x_2=-0.25$ and $p_2=0$. The continuous black lines in (a-b) guide the eye for slope $-1$. The continuous, dashed and dashed-dotted lines in (c-d) indicate slopes  $-\lambda_1/\ln\left(10\right)$, $-2\lambda_1/\ln\left(10\right)$ and $-4\lambda_1/\ln\left(10\right)$ respectively, with $\lambda_1 = 0.045$. 
    }
    \label{fig:C1_MLE_GALI_Rep_01}
\end{figure}

We also perform a similar analysis for a chaotic orbit of the HH system, with initial conditions  on the $x_1=0$ PSS of Fig.~\ref{fig:pss_hh_h=0.125}  $x_2=-0.25$, $p_2 = 0$, while,  once again, $p_1>0$ is selected such that $\mathcal{H}=0.125$. In Figs.~\ref{fig:C1_MLE_GALI_Rep_01}(a-b) we plot the time evolution of the two largest finite-time LEs computed using both the VM and MPM respectively. As time evolves, the computed $\lambda_1$ (blue curves) for both methods show a striking resemblance, saturating to values $\lambda_1 \approx 0.045$ at large time. On the other hand, the computed second LE (green curves) show a power law decay $\lambda_2 \propto t^{-1}$, which is denoted by the black lines in both panels.

For this chaotic orbit, we also computed the temporal dependence of the associated GALIs. In Figs.~\ref{fig:C1_MLE_GALI_Rep_01}(c-d) we show the time evolution of the $\mbox{GALI}_2$ (blue curves), $\mbox{GALI}_3$ (green curves) and $\mbox{GALI}_4$ (red curves). The numerical results of all GALIs for the VM are in accordance to  the theoretical expectations [Eq.~\eqref{eq:behavior_gali_chaotic_general}], as the values of $\mbox{GALI}_2$, $\mbox{GALI}_3$ and $\mbox{GALI}_4$ follow exponential decays with exponents $-\lambda_1/\ln\left(10\right) \approx -0.02$, $-2\lambda_1/\ln\left(10\right)\approx -0.04$ and $-4\lambda_1/\ln\left(10\right)  \approx -0.08$ respectively, for $\lambda_1 = 0.045$, which are indicated  by the continuous, dashed and dashed-dotted black lines in Fig.~\ref{fig:C1_MLE_GALI_Rep_01}(c) [these exponential decays are also indicated in the same way in Fig.~\ref{fig:C1_MLE_GALI_Rep_01}(d)]. The outcomes of the same computations using the MPM are presented in Fig.~\ref{fig:C1_MLE_GALI_Rep_01}(d). Overall, we see a good agreement between the VM and the MPM for the numerical values of the $\mbox{GALI}_2$, $\mbox{GALI}_3$ and $\mbox{GALI}_4$ for times $t\lesssim 200$. Nevertheless, for $t\gtrsim 200$ a divergence from the theoretical power laws can be seen with $\mbox{GALI}_2$ practically saturating at times  $t\gtrsim 500$.

\begin{figure*}[h!]
    \centering
    \includegraphics[width=\textwidth]{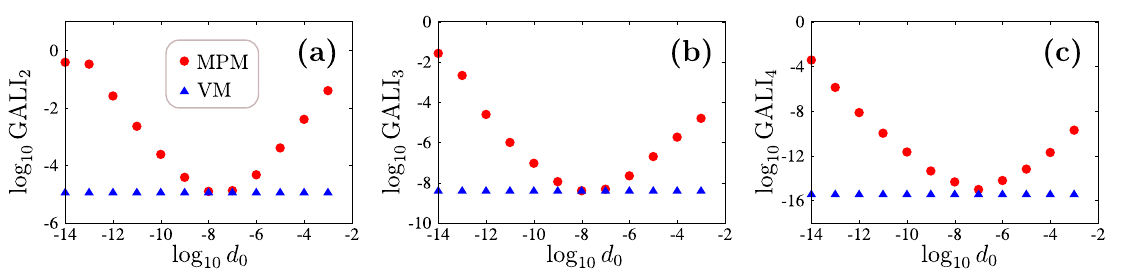}
    \caption{
        {\bf 2D HH system}.
        Dependence of the GALIs on the norm $d_0$ of the initial deviation vectors. Practical reliable regions of the computation of the (a) GALI$_2$, (b) GALI$_3$ and (c) GALI$_4$ for the HH model using the VM (blue triangles) and the MPM (red circles). The presented GALI values have been averaged  over $1000$ 
        sets of initial orthogonal tangent vectors of norm $d_0=\lVert \bm{w}_0\rVert = \lVert \bm{\delta}_0\rVert$ about the reference chaotic orbit with coordinates $x_1=0$, $x_2 = -0.25$, $p_1 =0.42$ and $p_2 = 0$. The final time of integration is $t= 200$ time units.
    }
    \label{fig:C1_dependence_GALI_vs_d0}
\end{figure*}

We now estimate the practical reliable regions of the computation of the GALIs by exploring the indices' dependence on the magnitude of separation vectors $d_0$. In Fig.~\ref{fig:C1_dependence_GALI_vs_d0}, we plot the values of the $\mbox{GALI}_2$, $\mbox{GALI}_3$ and $\mbox{GALI}_4$ at $t= 200$ for the chaotic orbit considered in Fig.~\ref{fig:C1_MLE_GALI_Rep_01}, as a function of $d_0$. We note that the presented results are obtained by averaging over $1000$ random sets of initial tangent vectors in order to mitigate the dependence of the computed GALIs on the choice of the initial deviation vectors, (see e.g.~Sect.~5.4.1.1 of \cite{SM2016}). The values of the GALIs computed using the VM are indicated by blue triangles, while the values obtained through the MPM are shown by red circles. The numerically computed $\mbox{GALI}_2$, $\mbox{GALI}_3$ and $\mbox{GALI}_4$ values using the VM are clearly independent of the magnitude of the initial deviation vectors in Fig.~\ref{fig:C1_dependence_GALI_vs_d0}. For the MPM however, a non-trivial relation between the $\mbox{GALI}_2$, $\mbox{GALI}_3$,  $\mbox{GALI}_4$ values and  $d_0$ is seen. Our results show $V-$like shapes with a global minimum value located at $d_0\approx 10^{-8}$ for all  presented GALIs. This result is in agreement with the theoretical prediction [see  Fig.~\ref{fig:GALI_ft_do_theoretical predictions}(a)] which established the global minimum at $d_0\approx \varepsilon^{1/2}$ in a computer environment with precision $\varepsilon$. Consequently, in the case of double precision computations for which  $\varepsilon\approx 10^{-16}$ the minimum occurs at $d_0\approx 10^{-8}$.

\begin{figure*}[h!]
    \centering
    \includegraphics[width=\textwidth, height=0.22\linewidth]{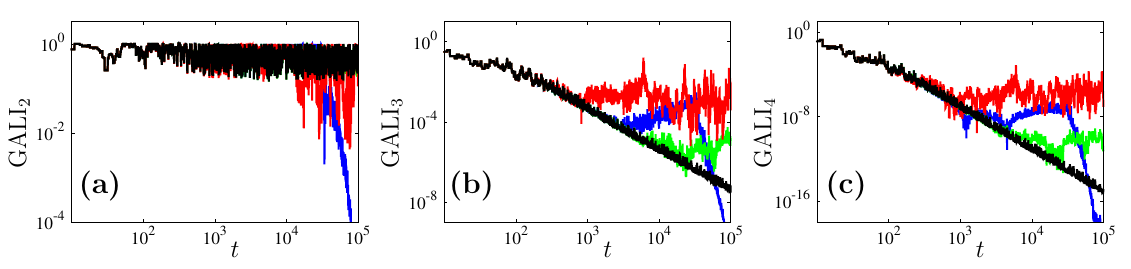}
    \caption{
       {\bf 2D HH system}.
        Time evolution of the  (a) $\mbox{GALI}_2$, (b) $\mbox{GALI}_3$ and (c) $\mbox{GALI}_4$ for the regular orbit of Fig.~\ref{fig:R1_MLE_GALI_Rep_01}. For the application of the MPM we use three different values for the initial norms of the deviation vectors: $d_0 = 10^{-5}$, $10^{-8}$ and $10^{-11}$ (respectively blue, green and red colored curves). The black curve in each panel represents  results obtained using the VM.
    }
    \label{fig:R1_dependence_GALI_vs_d0}
\end{figure*}

To further probe these findings on the numerical reliable regions, we numerically evaluate the GALIs using three different choices of the norm of the deviation vector $d_0$ for the regular orbit of Fig.~\ref{fig:R1_MLE_GALI_Rep_01}. In  Fig.~\ref{fig:R1_dependence_GALI_vs_d0}, we show the time evolution of the  $\mbox{GALI}_2$, $\mbox{GALI}_3$ and $\mbox{GALI}_4$ when computed using the MPM with initial norms of the separation vectors $d_0 = 10^{-5}$, $10^{-8}$ and $10^{-11}$ (blue, green and red curves respectively), and with the VM (black curves).

From the results of Fig.~\ref{fig:R1_dependence_GALI_vs_d0} we see that, in general, the computation of the GALIs for regular orbits is reliable for short-time evolution of  fairly small initial norms of the tangent vectors. Indeed, for all initial magnitudes of the separation vectors $d_0$ used in Fig.~\ref{fig:R1_dependence_GALI_vs_d0}, the computations of the $\mbox{GALI}_2$ [Fig.~\ref{fig:R1_dependence_GALI_vs_d0}(a)], $\mbox{GALI}_3$ [Fig.~\ref{fig:R1_dependence_GALI_vs_d0}(b)] and $\mbox{GALI}_4$ [Fig.~\ref{fig:R1_dependence_GALI_vs_d0}(c)] using the MPM practically overlap with the ones obtained using the VM up to $t\approx10^{3}$. For larger values of the integration time ($t\gtrsim 10^{3}$) we expect the accumulation of numerical errors in the case of the MPM to result in values of the computed GALIs which deviate from those obtained using the VM. Thus, based on the practical reliable regions of the MPM in Fig.~\ref{fig:C1_dependence_GALI_vs_d0}, we expect the rate of accumulation of numerical errors to be the smallest for the case with $d_0= 10^{-8}$. This is indeed what we observe in Fig.~\ref{fig:R1_dependence_GALI_vs_d0}. The values of the $\mbox{GALI}_2$, $\mbox{GALI}_3$ and $\mbox{GALI}_4$ computed with $d_0=10^{-8}$ [green curves in Fig.~\ref{fig:R1_dependence_GALI_vs_d0}] overlap with the results obtained with the VM for  longer integration times compared to the ones obtained using $d_0=10^{-5}$ and $d_0=10^{-11}$ [respectively blue and red curves in Fig.~\ref{fig:R1_dependence_GALI_vs_d0}].

\begin{figure}[h!]
    \centering    \includegraphics[width=\columnwidth]{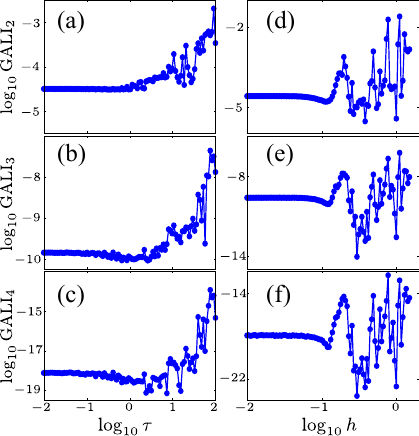}
    \caption{
    {\bf 2D HH system}. Dependence of the GALI$_2$, GALI$_3$ and GALI$_4$ on (a-c) the renormalization time $\tau$ and (d-f) the integration time step $h$ for the chaotic orbit  considered in Figs.~\ref{fig:C1_MLE_GALI_Rep_01} and~\ref{fig:C1_dependence_GALI_vs_d0}. In (a-c)  the integration time step is set to $h=0.01$, while in (d-f)  the renormalization time is fixed to $\tau = 1$. For all computations the final time of integration is $t=200$, and random initial deviation vectors of norm $d_0=10^{-9}$ were used. The GALI values are obtained through averaging  over $1000$ sets of random initial orthogonal deviation vectors. 
    }
    \label{fig:GALI_vs_trenorm_01a}
\end{figure}

Let us now investigate the impact of the renormalization time $\tau$, on the computation of the GALIs using the MPM. In order to do that, we focus on the chaotic orbit considered in Figs.~\ref{fig:C1_MLE_GALI_Rep_01} and~\ref{fig:C1_dependence_GALI_vs_d0}. For our computations we fix $d_0 = 10^{-9}$  and evolve the chaotic orbit and its nearby orbits up to $t= 200$ with using as integration time step $h=0.01$. We record the final values of the $\mbox{GALI}_2$ in Fig.~\ref{fig:GALI_vs_trenorm_01a}(a), the $\mbox{GALI}_3$ in Fig.~\ref{fig:GALI_vs_trenorm_01a}(b) and the $\mbox{GALI}_4$ in Fig.~\ref{fig:GALI_vs_trenorm_01a}(c), averaged over $1000$ sets of  random initial orthogonal separation vectors, varying $\tau$ in the interval between $0.01$ to $100$. We find that the dependence of the computed GALIs on the renormalization time $\tau$ can be split into two main regions. The first one corresponds to $\tau \lesssim 1$ for which the final values of the computed GALIs remain practically constant at around $5\times 10^{-5}$, $2\times 10^{-10}$ and $1.2\times 10^{-18}$ for the $\mbox{GALI}_2$, $\mbox{GALI}_3$ and $\mbox{GALI}_4$  respectively. On the other hand, in the region where $\tau\gtrsim 1$, we see that increasing the values of $\tau$ leads to a roughening and an overall increase in the final values  of the $\mbox{GALI}_2$, $\mbox{GALI}_3$ and $\mbox{GALI}_4$. This results hints toward the appearance of large numerical errors in computing the GALI values. 

\begin{figure}[t]
    \centering    \includegraphics[width=\columnwidth]{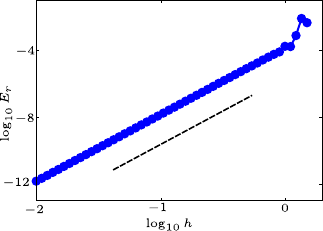}
    \caption{
    {\bf 2D HH system}.
    Dependence of the upper bound of the time evolution of the relative energy error $E_r$ \eqref{eq:relative_energy_error} on the integration time step $h$, for the chaotic orbit of Fig.~\ref{fig:C1_MLE_GALI_Rep_01}, when it is integrated up to  $t=200$ by the $ABA864$ symplectic integrator  of order $4$. The dashed black line indicates slope $4$.
    }
    \label{fig:er_vs_h_C1_hh}
\end{figure}

A similar behavior is also observed when computing the dependence of the computed GALIs on the integration time step $h$. For these numerical simulations, we kept the same set up for the numerical integration as above, setting $\tau = 1$ and  varying the integration time step $h$. The obtained results are depicted in Figs.~\ref{fig:GALI_vs_trenorm_01a}(d)-(f) for the $\mbox{GALI}_2$, $\mbox{GALI}_3$ and $\mbox{GALI}_4$ respectively. We see that the dependence of the final value of the GALIs on $h$ can also be split into two main regions, separated by $h\approx 10^{-1}$. For $h\lesssim0.1$, these GALI values are constant for all $h$, while for  $h\gtrsim0.1$ we obtain different final values of the GALIs for each time step $h$. The striking result here is that in terms of the relative energy error, $h=0.1$ corresponds to $E_r \approx 10^{-8}$ [Fig.~\ref{fig:er_vs_h_C1_hh}]. Thus, we conclude that numerical simulations with relative energy errors $E_r \gtrsim 10^{-8}$ are in general inaccurate. Such a precision could be easily achieved using high order numerical integration schemes. It is worth pointing out the slope of the $E_r$ against $h$ in Fig.~\ref{fig:er_vs_h_C1_hh} is $4$ (this slope is indicated by the dashed black line),  matching the order of the $ABA864$ symplectic integrator [see also Sect.~\ref{sec:model_descriptions}]. 

\begin{figure*}[t]
    \centering
    \includegraphics[width=\textwidth]{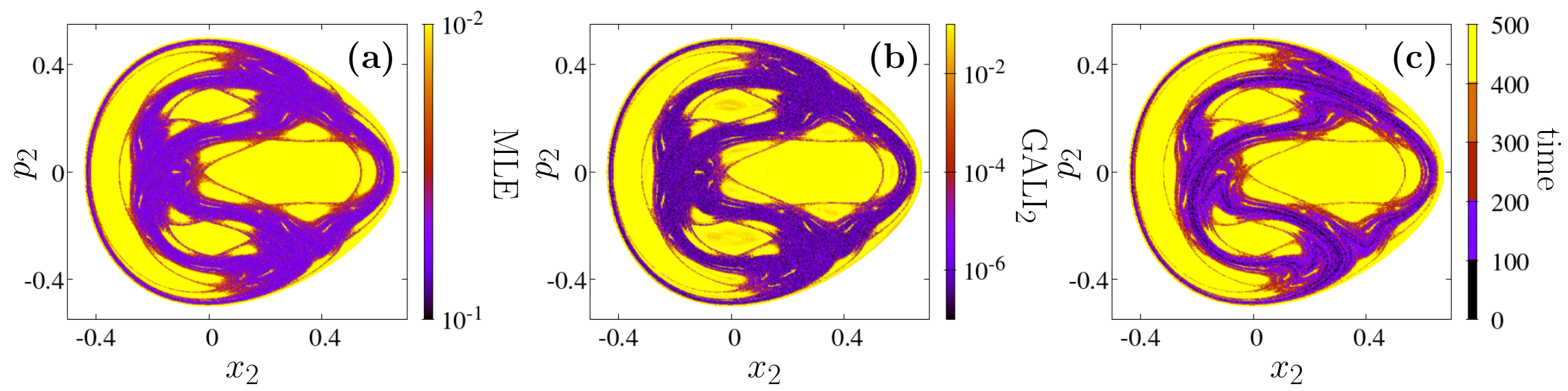}
    \caption{
    {\bf 2D HH system}.
    Regions of different dynamical behavior on the $x_1=0$, $p_1>0$ PSS of the HH system \eqref{eq:hamiltonian_hh}  with $\mathcal{H}=0.125$, identified  through (a) the MLE values at $t = 1000$, and (b) the $\mbox{GALI}_2$ values at $t = 500$. 
    (c) Regions of different values of the time $t$ needed for the $\mbox{GALI}_3$ to become less than $10^{-8}$ for the same set of initial conditions as in panels (a) and (b).
    In all computations $h=0.01$, $\tau = 1$ and $d_0=10^{-8}$. 
    In each panel the average GALI values over results obtain for $50$ sets of random initial orthogonal separation vectors are shown, with initial conditions colored according to the color scale at the right side of the panel. 
       }
    \label{fig:colormap_pss_h0.125_galis}
\end{figure*}

What about exploring the global dynamics of a dynamical system using the GALI method with the MPM? In Fig.~\ref{fig:colormap_pss_h0.125_galis}, we show colormaps of the PSS $x_1 = 0$ of Fig.~\ref{fig:pss_hh_h=0.125}, made of $700\times 600$ grid points. The $p_1$ coordinate of each initial condition on the ($x_2, p_2$) phase plane of Fig.~\ref{fig:colormap_pss_h0.125_galis} is set such that $p_1>0$ with $\mathcal{H} = 0.125$. The initial condition of each generated orbit on the $(x_2, p_2)$ PSS is colored according its MLE value at $t = 1000$ [Fig.~\ref{fig:colormap_pss_h0.125_galis}(a)] and its $\mbox{GALI}_2$ value at $t = 500$ [Fig.~\ref{fig:colormap_pss_h0.125_galis}(b)]. We also note that the presented results are obtained by averaging the indices values over $50$ configurations of sets of random orthogonal initial tangent vectors in order to mitigate their dependence on the choice of the initial conditions for the tangent vectors. Both chaos indicators display the same qualitative dynamical behavior in terms of characterizing the stability of orbits in the phase space. More specifically, Figs.~\ref{fig:colormap_pss_h0.125_galis}(a-b), show the same phase plane structure, that is to say the regular regions (yellow colored points)  and chaotic seas (purple colored points) are practically the same in both panels, and similar to what is revealed by the PSS of Fig.~\ref{fig:pss_hh_h=0.125}. In particular, we see that several small regular islands in the chaotic regions are well capture by the MPM, along with the so-called `sticky orbits' at the interfaces of regular islands with chaotic seas corresponding to a mixed blend of colors (e.g., yellow purple).  The difference between the two numerical simulations resides on the computational time. Indeed, it took around $t=35.7$ and $t=28.2$ CPU hours respectively for the simulations of Fig.~\ref{fig:colormap_pss_h0.125_galis}(a) and Fig.~\ref{fig:colormap_pss_h0.125_galis}(b) to be completed.

To the question whether we can reach the same accuracy analyzing the phase space of the HH system using GALIs of higher orders, the answer is yes. First we remind the reader that the GALIs of higher orders decrease for both regular and chaotic orbits. Thus, we need to follow a different approach than recording the final value of the GALIs as was done for GALI$_2$. In order to characterize the global dynamics of the HH system, we register the time it took the higher order GALIs to reach a particular small threshold value~\cite{SM2016}, exploiting the fact that the GALIs    decay faster for chaotic orbits compared to regular ones. In Fig.~\ref{fig:colormap_pss_h0.125_galis}(c), we color initial conditions according to the time it took $\mbox{GALI}_3$ to reach the threshold $10^{-8}$. This time is again averaged over $50$ different sets of initial orthogonal tangent vectors. Clearly large decay times close to the maximum considered time ($t=500$) are related to regular islands (yellow colored regions), while chaotic seas (purple colored regions) have in general smaller decay times ($t=100-200$). In addition, we also find initial conditions with intermediate values of decay times, which are in general at the edges of regular islands and belong to `sticky' chaotic orbits. Although we are now integrating more tangent vectors than for obtaining the results of Figs.~\ref{fig:colormap_pss_h0.125_galis}(a) and (b), the advantage of this approach is that we can stop the evolution of each orbit when the GALI$_3$ has reached the threshold. Following this methodology, we required  a computational time of $t=28.6$ CPU hours, close to what was obtained for the $\mbox{GALI}_2$ computations, despite having to integrate an additional  tangent vector for $\mbox{GALI}_3$. Thus, all these results  not only confirm the ability of the GALI method implemented using the MPM to characterize the global dynamics of Hamiltonian systems, but also demonstrate its computational speed and efficiency compared to the MLE approach.

\subsection{The $\beta$-Fermi-Pasta-Ulam-Tsingou model}
\label{subsec:fput_num}

We now explore the performances of the GALI method using another well-known Hamiltonian model: the $\beta$-FPUT lattice [Eq.~\eqref{eq:bFPUT}]. We first look at a small lattice, consisting of five ($N=5$) oscillators, to establish the basic behaviours of the orbits we are going to study using the MLE. Knowing from the linear stability analysis performed in \cite{AB2006} that the value of the first energy destabilization threshold is $\mathcal{H}_5^{c} = 7.4$ for the SPO1 with $N=5$, we choose (see \ref{app:sec:fputIC}) orbits whose  initial conditions are perturbations of the SPO1s at fixed energy with $\mathcal{H}_5=5$ [Eq.~\eqref{eq:IC_bFPUT_regular}]  and $\mathcal{H}_5 = 10$ [Eq.~\eqref{eq:IC_bFPUT_chaotic}] respectively below and above $\mathcal{H}_5^{c}$. For the integration of the equations of motion and the related variational equations, we fix the integration time step to $h=0.02$, which results in the relative energy error \eqref{eq:relative_energy_error} being always bounded from above by $E_r^N \approx 10^{-11}$. In addition, for both the VM and the MPM we set the renormalization time to $\tau=1$.

\begin{figure*}[ht!]    
\centering
\includegraphics[width=0.8\textwidth]{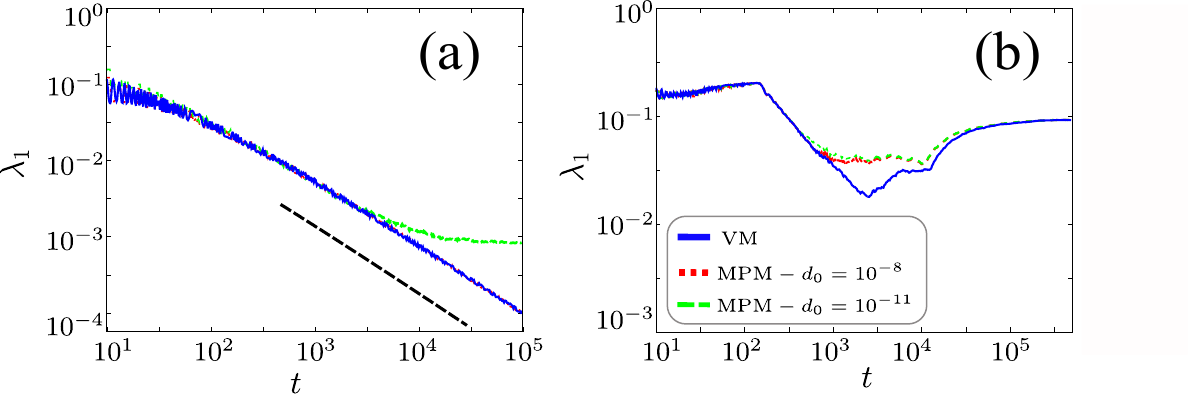}
    \caption{
    {\bf $\bm{\beta}$-FPUT chain.}
    The time evolution of the ftmLE $\lambda_1$ for two perturbations of the SPO1 of the $\beta$-FPUT model with $N=5$ oscillators: (a) A regular orbit for energy $\mathcal{H}_5=5$, and (b) a chaotic orbit with energy $\mathcal{H}_5=10$, whose initial conditions are given in Eq.~\eqref{eq:IC_bFPUT_regular}  and Eq.~\eqref{eq:IC_bFPUT_chaotic} respectively. Note that in both panels the solid blue curves correspond to results computed using the VM, while red dotted and green dashed curves  show  results obtained by the MPM for an initial random  deviation vector of norm $d_0=10^{-8}$ and $d_0=10^{-11}$ respectively. The dashed black line in (a) guides the eye for slope $-1$.
    }
    \label{fig:mle_ft_time_bfput}
\end{figure*}

As a representative case, the time evolution of the MLE for the regular with $\mathcal{H}_5=5$ is presented in Fig.~\ref{fig:mle_ft_time_bfput}(a), computed using the VM (continuous blue curve) as well as the MPM. In the case of the MPM, we use an initial  separation vector of magnitude  $d_0 = 10^{-8}$ (dotted red curve) and $d_0=10^{-11}$ (dashed green curve). As Fig.~\ref{fig:mle_ft_time_bfput}(a) shows, the MLE values decay towards zero  following the power law $t^{-1}$ (shown by the black dashed line) for the VM, confirming the regular nature of the orbit we are studying. Using exactly the same  separation vectors of the VM for the MPM calculations,  rescaled to have norms $d_0=10^{-8}$ and $d_0=10^{-11}$, we see that we manage to  accurately track the VM result until the last stage of the evolution when  $d_0 = 10^{-8}$, while  the results for $d_0  = 10^{-11}$ start deviating from the ones obtained by the VM for $t \gtrsim 10^{3.5}$. On the other hand,  for the chaotic orbit with $\mathcal{H}_5 = 10$, we  see in Fig.~\ref{fig:mle_ft_time_bfput}(b) that all simulations show similar time evolution of the MLE. In particular, the computed indices eventually converge to practically the same positive, constant value. This behavior clearly establishes the chaotic nature of the orbit. It is worth noting that the explanation for the inaccuracy of the MPM computations of the MLE for $d_0 = 10^{-11}$ can be found in Ref.~\cite{MH2018}, and is related  to the limitations of the MPM when the norms of the separation vectors are  too close to the machine precision. 

\begin{figure*}[hbt]
    \centering 
    \includegraphics[width=0.7\textwidth]{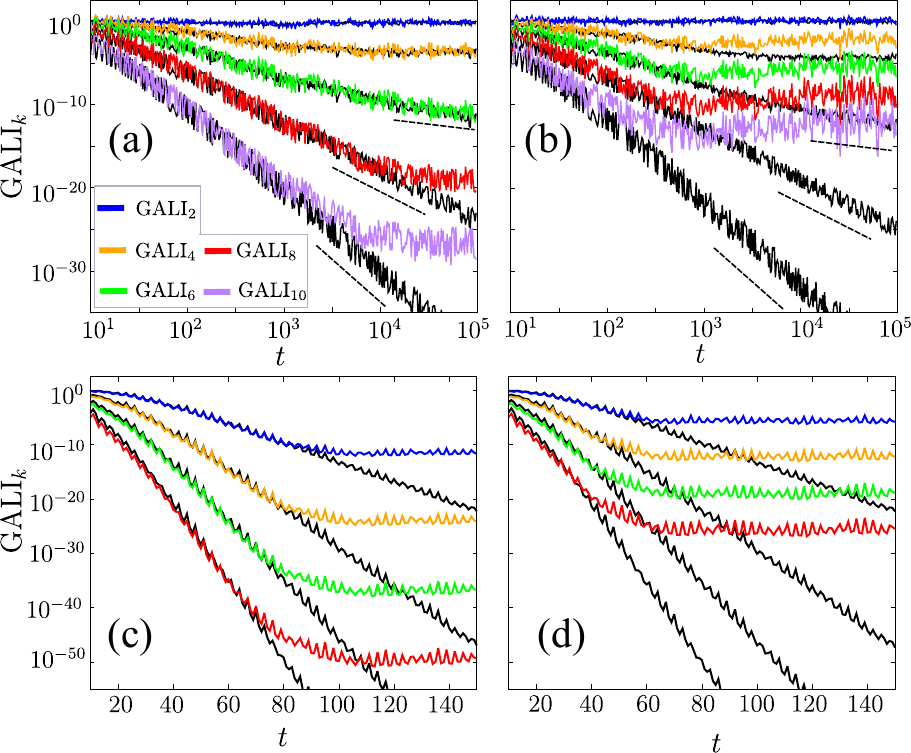}
    \caption{
    {\bf $\bm{\beta}$-FPUT chain.}
    Time evolution of several computed GALIs  for the regular [(a) and (b)] and the chaotic [(c) and (d)] orbits of Fig.~\ref{fig:mle_ft_time_bfput}. In all panels black curves correspond to results obtained by the VM, while colored curves show the results of the MPM. In (a) and (c) deviation vectors of norm $10^{-8}$ were used, while for (b) and (d) a norm of $10^{-11}$ was considered.
    The  grey dashed lines in (a) and (b) show the theoretically predicted slopes for the various GALI$_k$  with $k>N$, according to Eq.~\eqref{eq:behavior_gali_regular_general}.
    Note that in  (a) and (b) time (the horizontal axis) is in logarithmic scales, while in (c) and (d) in linear scales.
    }
    \label{fig:gali_ft_time_bfput}
\end{figure*}

We also study the behavior of the various GALIs for these two orbits of the $\beta$-FPUT model. In Figs.~\ref{fig:gali_ft_time_bfput}(a) and (b) the black curves  from the top to the bottom respectively depict the $\mbox{GALI}_2$, $\mbox{GALI}_4$, $\mbox{GALI}_6$, $\mbox{GALI}_8$ and $\mbox{GALI}_{10}$ values calculated by the VM as a funtion of time $t$ of  the regular orbit of Fig.~\ref{fig:mle_ft_time_bfput}(a). Similarly to the HH system, the behavior of the VM-calculated GALI values follow the theoretical predictions, irrespective of the choice of the initial norm  of the tangent vectors, namely $d_0=10^{-8}$ [Fig.~\ref{fig:gali_ft_time_bfput}(a)] and $d_0=10^{-11}$ [Fig.~\ref{fig:gali_ft_time_bfput}(b)]. Thus, we see that  the $\mbox{GALI}_2$ and $\mbox{GALI}_4$ values tend to saturate toward constant positive values, while  the computed $\mbox{GALI}_6$,  $\mbox{GALI}_8$ and $\mbox{GALI}_{10}$ decay following the power laws of Eq.~\eqref{eq:behavior_gali_regular_general} (which are depicted by dashed gray lines) until  the final time of our simulations. On the other hand, the results obtained by the MPM for $\mbox{GALI}_2$ (blue curve), $\mbox{GALI}_4$ (orange curves), $\mbox{GALI}_6$ [(green curves), $\mbox{GALI}_8$ (red curves) and $\mbox{GALI}_{10}$ (purple curves) eventually show deviations from the theoretically expected behaviors. In particular, we  find that the MPM-calculated GALI values $d_0=10^{-8}$ [Fig.~\ref{fig:gali_ft_time_bfput}(a)] practically coincide with the ones computed using the VM, for times up to $t\approx 10^4$ for the $\mbox{GALI}_{8}$ and $\mbox{GALI}_{10}$, while  the results for the  $\mbox{GALI}_2$, $\mbox{GALI}_4$ and $\mbox{GALI}_6$ are similar to the ones of the VM, until our final computation time. However, this compatibility between the two methods lasts for a shorter time ($t\approx 10^{3}$) in the case of the computation using $d_0=10^{-11}$ in Fig.~\ref{fig:gali_ft_time_bfput}(b), with only GALI$_2$ results being similar to the VM ones until the end of our computations. It is worth noting that in both cases (i.e., for $d_0=10^{-8}$ and $d_0=10^{-11}$) the computation of the $\mbox{GALI}_2$ remains rather robust since both methods return practically the same results.

The GALI values of the chaotic orbit shown in Fig.~\ref{fig:mle_ft_time_bfput}(b) should decay to zero exponentially fast according to Eq.~\eqref{eq:behavior_gali_chaotic_general}. This is exactly what is seen in time evolution of the $\mbox{GALI}_2$, $\mbox{GALI}_4$, $\mbox{GALI}_6$, $\mbox{GALI}_8$ and $\mbox{GALI}_{10}$ values [black curves from top to bottom respectively in Figs.~\ref{fig:gali_ft_time_bfput}(c) and (d)] computed using the VM with $d_0 = 10^{-8}$ [Fig.~\ref{fig:gali_ft_time_bfput}(c)] and $d_0=10^{-11}$ [Fig.~\ref{fig:gali_ft_time_bfput}(d)].On the other hand, the values obtained by the  MPM tend again to follow similar temporal behaviors to the ones of the VM until $t\approx 80$ [Fig.~\ref{fig:gali_ft_time_bfput}(c)] and $t\approx 50$ [Fig.~\ref{fig:gali_ft_time_bfput}(d)]. After these times, a clear deviation of the results of the MPM from those of the VM is observed  for all GALIs.

\begin{figure*}[htb!]
    \centering 
    \includegraphics[width=0.7\textwidth]{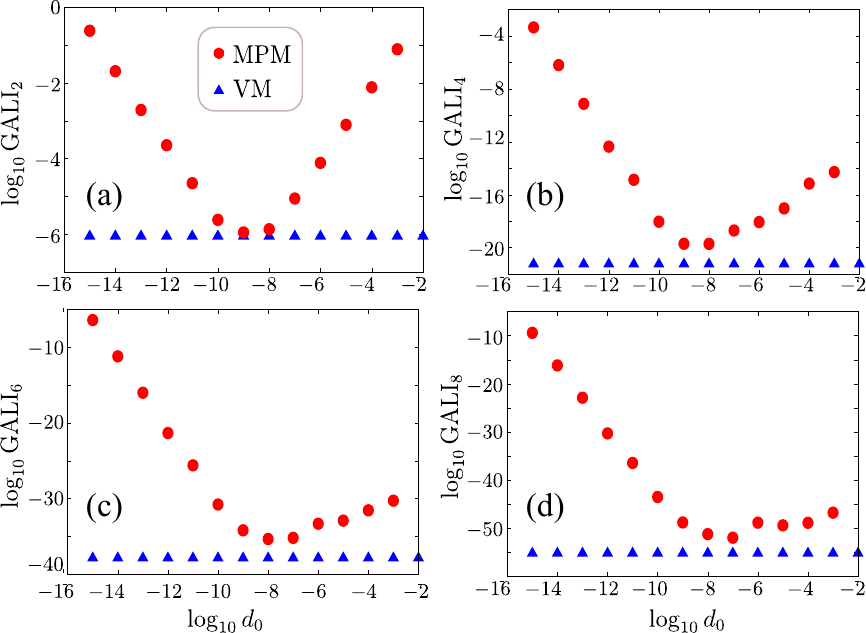}
    \caption{
    {\bf $\bm{\beta}$-FPUT chain.} Reliable regions for the computation of the (a) GALI$_2$, (b) GALI$_4$, (c)GALI$_6$, and (d) GALI$_8$ of the chaotic orbit of the $\beta$-FPUT lattice considered in Fig.~\ref{fig:mle_ft_time_bfput}. using the VM (blue triangles) and the MPM (red circles). The presented GALI values have been obtained by averaging over $100$ sets of initial orthogonal tangent vectors. The final time of integration is $t= 200$ time units. 
    }
    \label{fig:gali_ft_do_ft_N5_bfput}
\end{figure*}

Let us now check  the reliable region  for the GALIs computed by the MPM, considering again the chaotic orbit of Fig.~\ref{fig:mle_ft_time_bfput}(b). In our simulations we fix the integration time step to $h=0.02$ and evolve the required orbits  up to $t= 200$ time units. 
We also  set the renormalization time to $\tau=1$ and vary the $d_0$ values in the interval $[10^{-14},10^{-2}]$, with unit step in logarithmic scale. The GALI values at the end of the integration time are recorded, and averaged over $100$ sets of random initial orthogonal tangent vectors. The outcomes of this process are presented in Fig.~\ref{fig:gali_ft_do_ft_N5_bfput}, where results for both the VM (blue filled-triangles) and the MPM  (red filled-circles) are shown for $\mbox{GALI}_2$ [Fig.~\ref{fig:gali_ft_do_ft_N5_bfput}(a)], $\mbox{GALI}_4$ [Fig.~\ref{fig:gali_ft_do_ft_N5_bfput}(b)], $\mbox{GALI}_6$ [Fig.~\ref{fig:gali_ft_do_ft_N5_bfput}(c)], and $\mbox{GALI}_8$ [Fig.~\ref{fig:gali_ft_do_ft_N5_bfput}(d)]. Clearly the numerical uncertainties of these GALIs  do not depend on the value of $d_0$ for the VM as their values remain constant for all considered $d_0$. On the other hand, for the results of the MPM  we observe an overall $V-$shape dependence of the GALI values against $d_0$ with the minimum located at $d_0 \approx 10^{-8}$ for all cases. This behavior is again in agreement with the prediction of Eq.~\eqref{eq:gali_two_particle_equation_00}.

It is worth commenting a bit more on the $V-$shapes of Fig.~\ref{fig:gali_ft_do_ft_N5_bfput}, which appear to be more symmetric for small GALI orders [e.g.~$\mbox{GALI}_2$ in Fig.~\ref{fig:gali_ft_do_ft_N5_bfput}(a)] compared to larger ones [e.g.~$\mbox{GALI}_8$ in Fig.~\ref{fig:gali_ft_do_ft_N5_bfput}(d)] with respect to the vertical line passing through the minimum. Although we do not show these results, in order to avoid repetitions of similarly looking plots, analogous, slightly tilted  $V-$shapes were also obtained for the HH model as the time at which results were  computed increased from $t= 200$ [Fig.~\ref{fig:C1_dependence_GALI_vs_d0}(b)] to $t= 500$ for $\mbox{GALI}_3$, and from $t= 200$ [Fig.~\ref{fig:C1_dependence_GALI_vs_d0}(c)] to $t=300$ for $\mbox{GALI}_4$. We emphasize  that the symmetric nature of the $V-$shape of the reliable region of the GALI values is a consequence of the leading order term approximation of the numerical error [as was graphically represented in Fig.~\ref{fig:GALI_ft_do_theoretical predictions}]. This property holds true  for the $\mbox{GALI}_2$ computations. Consequently, the asymmetric $V-$shape in the results for the $\mbox{GALI}_4$  [Fig.~\ref{fig:gali_ft_do_ft_N5_bfput}(b)], $\mbox{GALI}_6$  [Fig.~\ref{fig:gali_ft_do_ft_N5_bfput}(c)] and $\mbox{GALI}_8$  [Fig.~\ref{fig:gali_ft_do_ft_N5_bfput}(d)] at time $t\approx 200$ indicates  that the leading order approximation of the error is no longer very accurate. In fact, at large times higher   order terms of the polynomial expansion of the numerical errors of the GALI computation should also start contributing meaningfully to the global error. Nevertheless, it is remarkable to see that even in cases beyond the validity of our first order approximation, our estimation about the optimal $d_0$ value (i.e., $d_0 \approx 10^{-8}$) remains accurate, something  which strongly supports the robustness of our findings.  It is worth emphasizing that we have found similarly looking  reliable regions for the GALI method computed using the MPM for other chaotic orbits of  $\beta$-FPUT models with various lattice sizes. Some of these results, in particular for $N=11$, $N=101$, and $N=501$ are shown in Fig.~\ref{fig:gali_ft_do_ft_N_bfput} of~\ref{app:sec:gali_for_varous_N_bFPUT}. 

\section{Conclusions and outlook}
\label{sec:conclusion_and_outlook}

We have shown that it is possible to accurately detect chaos in multidimensional Hamiltonian systems by  the GALI technique, using the MPM in place of the more commonly applied VM, which involves the system's variational equations that require the related equations of motion to be continuous and differentiable. We performed both a theoretical and a numerical analysis of the errors associated with the application of the MPM to evaluate the GALIs, and found that for relatively short evolution times the method's accuracy is comparable to that of the VM.

The numerical errors in the calculation of the GALI when utilizing the MPM depend primarily on the choice of the deviation vector magnitude $d_0$, the renormalization time $\tau$, and the global truncation error of the  numerical integration scheme $\varepsilon_T$ (Sect.~\ref{sec:theoretical_estimate}).
The latter error can be estimated by the largest value of the relative energy error $E_r$ in the case of Hamiltonian systems.

Testing these theoretical predictions for two well-known Hamiltonian models, namely the 2D HH system and the $\beta$-FPUT model for various values of its degrees of freedom, we confirmed in Sect.~\ref{sec:numerical_results} that, for sufficiently short times, whose duration depend on the order of the GALI (i.e., a higher GALI order results to smaller times), the MPM results match the ones obtained by the VM  when a reasonable deviation vector magnitude is chosen.
The choice of this magnitude was discussed and investigated in detail, along with its relation to the numerical limit of double precision accuracy ($\varepsilon \approx 10^{-16}$).
We found that an initial deviation vector norm of $d_0\approx10^{-8}$, a renormalization time $\tau \lesssim 1$ and a relative energy error $E_r \lesssim 10^{-8}$ are optimal in all cases.
Furthermore, we demonstrated that the MPM can be implemented to efficiently capture the global chaotic dynamics of the HH system, providing functionally equivalent results to the VM, something which constitutes a further confirmation of its validity for the practical computation of the GALI.

It is worth noting that the introduced MPM extends the idea of computing the most commonly used  chaos  indicator (i.e., the MLE) by considering two nearby  orbits instead of solving the system's variational equations to the case of the GALI chaos detection method. This is a computationally much harder task, 
as the evaluation the GALI of order $k$ by the MPM requires the simultaneous evolution of $k$ orbits in the neighborhood of the tested trajectory (along with the tested orbit itself), while for the  estimation of the MLE only one additional orbit is needed. Nevertheless, the GALI method is known to have many advantages over the computation of the MLE \cite{SBA2007,SBA2008,SM2016}. Thus, our work provides a way to exploit these advantages even when the VM cannot be used.

In the framework of the MPM, it will be interesting in the future to investigate the possible ability of  the short-term statistics of GALI values  to enhance the efficiency and speed of numerical computations allowing the early discrimination between regular and chaotic motion, a feature which could possibly facilitate the  integration of this approach with neural network systems~\cite{FJZWL2020}.

Our work paves the way for the exploration of chaotic dynamics using the GALI chaos detection method in a broader range of dynamical systems, including those where variational equations are difficult, or even impossible to derive analytically, or when their formulation introduces significant computational complexity. Such dynamical systems could be models whose equations of motion exhibit discontinuities, or systems having empirical, non-Newtonian, non-Hamiltonian and higher-dimensional field vectors. For instance, extending our investigation of the applicability and performance of the MPM to compute GALIs for the HH system, it would be interesting to study the characteristics chaos for the relativistic version of the system~\cite{PSM2023,SPM2024} using similar  approaches. Another interesting direction for applying the methodology developed in our work is to study chaos in pattern forming and turbulent far-from-equilibrium systems. Such studies could include  active fluids and solids whose equations of motion often extend the classical form including time-dependent or asymmetric stress/strain tensors.  Investigations of this kind could  be beneficial in novel applications like in swarm and soft robotic designs and bioengineering.

\section*{Acknowledgments}
We are grateful to Henok  Moges for useful discussions.
We also thank the Centre for High Performance Computing (CHPC) of South Africa for providing  computational resources for this project.
M.H.~acknowledges support from the National Research Foundation (NRF) of South Africa (Grant No. 129630).
The authors would like to express their gratitude to the anonymous reviewers for their comments and suggestions, which helped improve the clarity of this manuscript.

\appendix
\section{Pseudo-code for the computation of the generalized alignment index using the multi-particle method}
\label{app:sec:pseudo_code}

\noindent Inputs:
\begin{enumerate}
    \item Equations of motion of the system.
    \item Initial conditions of the tested orbit.
    \item The order $k$ of the computed GALI.
    \item Initial orthogonal separation vectors, $w_{i}(0)$, $i=1,2,\ldots, k$.
    \item Separation vector size $d_0$.
    \item Renormalization time $\tau$.
    \item Time step $\delta t$.
    \item Final integration time $t_f$.
\end{enumerate}
\hrule
\vspace{6pt}
\textit{Create initial conditions $xn_{i}$ for the nearby orbits from separation vectors.}\\
for i = 1 to k do\\
\indent $xn_i = x + w_i$\\
end\\
\textit{Set the first renormalization time.}\\
$T_\textit{next-renorm}=\tau$\\
\textit{Forward integration from time $t=0$ and computation of the GALI$_k$.}\\
while $t<t_f$ do \\
\indent $x(t)\rightarrow x(t+\delta t)$\\
\indent for i = 1 to k do\\
\indent\indent $xn_i(t) \rightarrow xn_i(t+\delta t)$\\
\indent end\\
\indent \textit{Check if the renormalization interval has passed}\\
\indent if $t>T_{\textit{next-renorm}}$ do\\
\indent\indent for i = 1 to k do\\
\indent\indent\indent $xn_i = x + (xn_i-x)\times d_0/||xn_i-x||$\\
\indent\indent end\\
\indent\indent $T_{\textit{next-renorm}} = T_{\textit{next-renorm}}+\tau$\\
\indent\indent \textit{Create matrix $\mathbf{A}$ having the unit deviation vectors as rows}\\
\indent\indent for i = 1 to k do\\
\indent\indent\indent $\mathbf{A}_i = (xn_i-x)/d_0$\\
\indent\indent end\\
\indent\indent \textit{Compute the singular values $z_i$ of $\mathbf{A}$}\\
\indent\indent $z_i$ = SVD($\mathbf{A}$)\\
\indent\indent for i = 1 to k do \\
\indent\indent\indent GALI$_{i}$ = $\prod_{j=1}^{k}z_j$\\
\indent\indent end\\
\indent end if\\
end
\vspace{6pt}
\hrule

\vspace{0.5cm}

In the above pseudo-code, the $\mbox{SVD}$ denotes the `single value decomposition' method, which has been used in \cite{SBA2008} to efficiently compute  GALIs.

\section{Initial conditions of orbits of the $\beta$-Fermi-Pasta-Ulam-Tsingou lattice with $N=5$}
\label{app:sec:fputIC}

The initial conditions of the orbits of the $\beta$-FPUT model \eqref{eq:bFPUT} with $N=5$ used to compute the MLE and the GALI in Figs.~\ref{fig:mle_ft_time_bfput}, \ref{fig:gali_ft_time_bfput} and \ref{fig:gali_ft_do_ft_N5_bfput} are 
\begin{equation}
    \begin{split}
          x_1 &= 1.03003,\\
          x_2 &= 0,\\
          x_3 &= -1.04003,\\
          x_4 &= 0,\\
          x_5 &= 1.04003,\\
          p_1 &= 0.29284,\\
          p_2 &= p_3 = p_4 = p_5 = 0,
    \end{split}
    \label{eq:IC_bFPUT_regular}
\end{equation}
for the regular orbit, and 
\begin{equation}
    \begin{split}
        x_1 &= -1.151344372237934177,\\
        x_2 &= -0.000000000000007443,\\
        x_3 &= 1.151344372237925295,\\
        x_4 &= 0.000000000000006056,\\
        x_5 &= -1.151344372237918634,\\
        p_1 &= 1.502757415151290132,\\
        p_2 &= 0.000000000000014786,\\
        p_3 &= -1.502757415151279696,\\
        p_4 &= -0.000000000000017280,\\
        p_5 &= 1.502757415151272147,        
    \end{split}
    \label{eq:IC_bFPUT_chaotic}
\end{equation}
for the chaotic one.

\section{Reliable regions of the generalized alignment index computed using the multi-particle method for various $\beta$-Fermi-Pasta-Ulam-Tsingou lattice sizes}
\label{app:sec:gali_for_varous_N_bFPUT}

\begin{figure*}[tbp!]
    \centering 
    \includegraphics[width=0.8\textwidth]{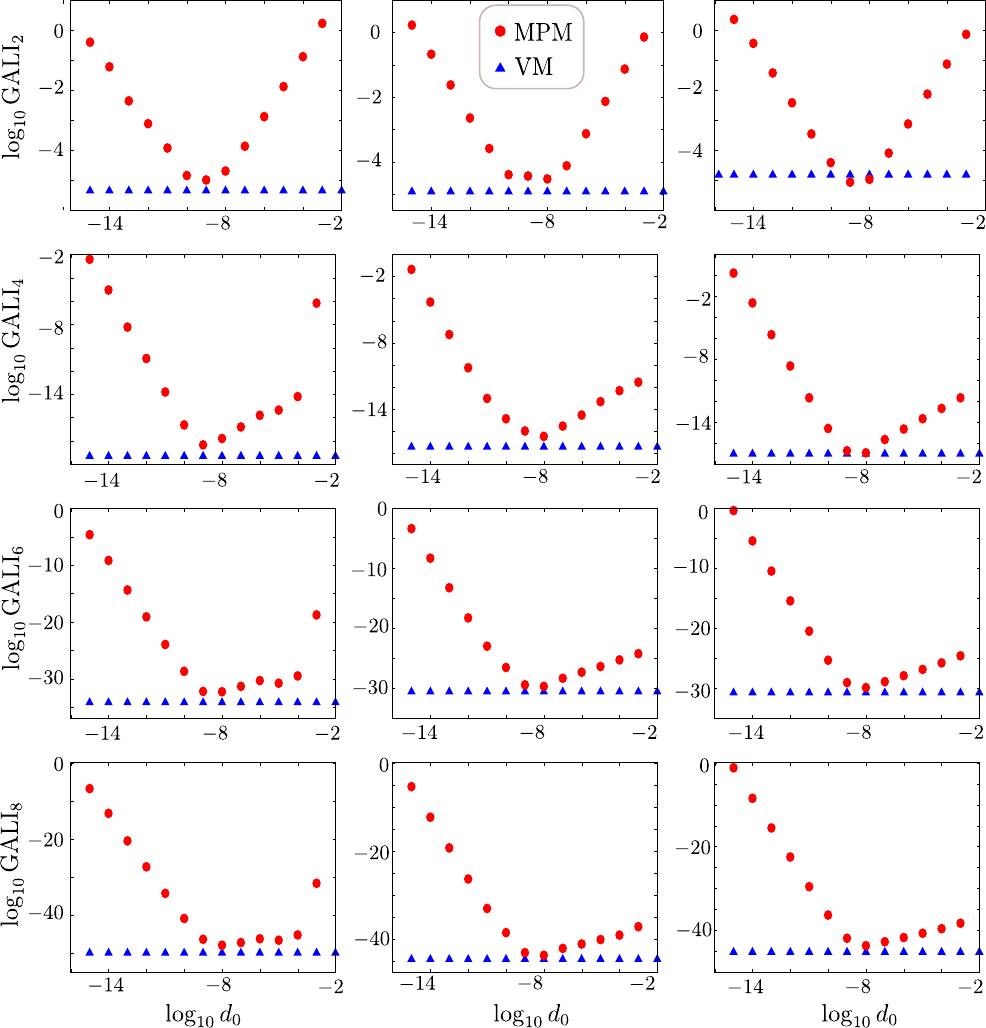}
    \caption{
    {\bf $\bm{\beta}$-FPUT chain.}
    Dependence of the GALI$_2$ (top row), GALI$_4$ (second row), GALI$_6$ (third row) and GALI$_8$ (bottom row) values on the norm $d_0$ of the initial deviation vectors for chaotic orbits of the $\beta$-FPUT lattice model \eqref{eq:bFPUT} $N=11$ (left column), $N=101$ (middle column), and $N=501$ (right column) oscillators, when the the MPM (red circles) and the VM (blue triangles) are used. In all cases the  norm $d_0$  is varied between $10^{-14}$ and $10^{-2}$. The GALI values are computed at time $t=700$ ($N=11$), $t=700$ ($N=101$), and $t=7000$ ($N=501$). The energies of the considered orbits are chosen just above the critical energies $\mathcal{H}_{11}^{c}=1.98$, $\mathcal{H}^{c}_{101}=1.515$ and $\mathcal{H}_{501}^{c}=1.502$ respectively for $N=11$, $101$ and $501$. 
     }
    \label{fig:gali_ft_do_ft_N_bfput}
\end{figure*}

In Sect.~\ref{subsec:fput_num} we presented results for the computation of the GALIs  using a $\beta$-FPUT lattice of $N=5$ oscillators. Here, for completeness sake, we discuss the reliable region of some GALI computations   for chaotic orbits of the $\beta$-FPUT model with  more degrees of freedom. The considered chaotic orbits are located in the neighborhood of the relevant unstable SPO1 trajectory for energies just above the first energy destabilization of the SPO1 happening at  $\mathcal{H}_{11}^{c}=1.98$, $\mathcal{H}^{c}_{101}=1.515$ and $\mathcal{H}_{501}^{c}=1.502$ respectively for $N=11$, $101$ and $501$. In Fig.~\ref{fig:gali_ft_do_ft_N_bfput} we present the final value of the GALI$_2$ (first, top row), GALI$_4$ (second row), GALI$_6$ (third row) and GALI$_8$ (fourth, bottom row)  computed by the MPM (red circles) and the VM (blue triangles) for  $N=11$ (left column), $N=101$ (middle column), and $N=501$ (right column) oscillators. Here we see again the same pattern which was observed for the $N=5$ [Fig.~\ref{fig:gali_ft_do_ft_N5_bfput}], with the optimal deviation vector magnitude being $d_0\approx 10^{-8}$ for all cases, along with the formation of a $V-$shaped feature, which becomes more asymmetric as the order of the GALI increases. These results confirm that the main findings of our theoretical analysis for the determination of the optimal $d_0$ value, are also valid for Hamiltonian systems with  many degrees of freedom.



\let\itshape\upshape
\bibliographystyle{elsarticle-num}
\bibliography{references.bib}

\begin{thebibliography}{10}
\expandafter\ifx\csname url\endcsname\relax
  \def\url#1{\texttt{#1}}\fi
\expandafter\ifx\csname urlprefix\endcsname\relax\def\urlprefix{URL }\fi
\expandafter\ifx\csname href\endcsname\relax
  \def\href#1#2{#2} \def\path#1{#1}\fi

\bibitem{H2000}
R.~C. Hilborn, Chaos and nonlinear dynamics: an introduction for scientists and
  engineers, Oxford University Press on Demand, 2000.

\bibitem{CARPINTERO201419}
D.~Carpintero, N.~Maffione, L.~Darriba, Lp-vicode: {A} program to compute a
  suite of variational chaos indicators, Astronomy and Computing 5 (2014) 19.

\bibitem{SGL2016}
C.~Skokos, G.~A. Gottwald, J.~Laskar, Chaos Detection and Predictability, Vol.
  915 of Lecture Notes in Physics, Springer, 2016.

\bibitem{PP2016}
A.~Pikovsky, A.~Politi, {L}yapunov exponents: a tool to explore complex
  dynamics, Cambridge University Press, 2016.

\bibitem{BGGS1980a}
G.~Benettin, L.~Galgani, A.~Giorgilli, J.-M. Strelcyn, {L}yapunov
  characteristic exponents for smooth dynamical systems and for {H}amiltonian
  systems; a method for computing all of them. {P}art 1: {T}heory, Meccanica 15
  (1980) 9.

\bibitem{BGGS1980b}
G.~Benettin, L.~Galgani, A.~Giorgilli, J.-M. Strelcyn, {L}yapunov
  characteristic exponents for smooth dynamical systems; a method for computing
  all of them. {P}art 2: {N}umerical application, Meccanica 15 (1980) 21.

\bibitem{S2010}
C.~Skokos, The {L}yapunov characteristic exponents and their computation,
  Lecture Notes in Physics 790 (2010) 63.

\bibitem{AKEDBK2018}
J.~Awrejcewicz, A.~V. Krysko, N.~P. Erofeev, V.~Dobriyan, M.~A. Barulina, V.~A.
  Krysko, Quantifying chaos by various computational methods. {P}art 1:
  {S}imple systems, Entropy 20 (2018) 175.

\bibitem{RG2020}
M.~Rautenberg, M.~G\"arttner, Classical and quantum chaos in a three-mode
  bosonic system, Phys. Rev. A 101 (2020) 053604.

\bibitem{S2024}
W.~Szumi{\'n}ski, A new model of variable-length coupled pendulums: from
  hyperchaos to superintegrability, Nonlinear Dynamics 112 (2024) 4117--4145.

\bibitem{SM2024}
W.~Szumi{\'n}ski, A.~J. Maciejewski, Dynamics and non-integrability of the
  double spring pendulum, Journal of Sound and Vibration 589 (2024) 118550.

\bibitem{FLG1997}
C.~Froeschl{\'e}, E.~Lega, R.~Gonczi, Fast {L}yapunov indicators. {A}pplication
  to asteroidal motion, Celestial Mechanics and Dynamical Astronomy 67 (1997)
  41.

\bibitem{LGF2016}
E.~Lega, M.~Guzzo, C.~Froeschl{\'e}, Theory and applications of the fast
  {L}yapunov indicator ({FLI}) method, Lecture Notes in Physics 915 (2016) 35.

\bibitem{CS2000}
{Cincotta, P. M.}, {Sim\'o, C.}, Simple tools to study global dynamics in
  non-axisymmetric galactic potentials - {I}, Astronomy and Astrophysics
  Supplement Series 147 (2000) 205.

\bibitem{CG2016}
P.~M. Cincotta, C.~M. Giordano, Theory and applications of the mean exponential
  growth factor of nearby orbits ({MEGNO}) method, Lecture Notes in Physics 915
  (2016) 93.

\bibitem{S2001}
C.~Skokos, Alignment indices: a new, simple method for determining the ordered
  or chaotic nature of orbits, Journal of Physics A: Mathematical and General
  34 (2001) 10029.

\bibitem{SBA2007}
C.~Skokos, T.~C. Bountis, C.~Antonopoulos, Geometrical properties of local
  dynamics in {Hamiltonian} systems: The generalized alignment index ({GALI})
  method, Physica D: Nonlinear Phenomena 231 (2007) 30.

\bibitem{SM2016}
C.~Skokos, T.~Manos, The smaller ({SALI}) and the generalized ({GALI})
  alignment indices: {E}fficient methods of chaos detection, Lecture Notes in
  Physics 915 (2016) 129.

\bibitem{LSB2019}
R.~J. Lewis-Swan, A.~Safavi-Naini, J.~J. Bollinger, A.~M. Rey, Unifying
  scrambling, thermalization and entanglement through measurement of fidelity
  out-of-time-order correlators in the {D}icke model, Nature Communications 10
  (2019) 1581.

\bibitem{HKSS2019}
M.~Hillebrand, G.~Kalosakas, A.~Schwellnus, C.~Skokos, Heterogeneity and chaos
  in the {P}eyrard-{B}ishop-{D}auxois {DNA} model, Physical Review E 99 (2019)
  022213.

\bibitem{ZDSS2020}
E.~E. Zotos, F.~L. Dubeibe, A.~F. Steklain, T.~Saeed, Orbit classification in a
  disk galaxy model with a pseudo-{N}ewtonian central black hole, Astronomy \&
  Astrophysics 643 (2020) {A}33.

\bibitem{KES2022}
E.~Kov\'ari, B.~\'Erdi, Z.~S\'andor, Application of the {S}hannon entropy in
  the planar (non-restricted) four-body problem: the long-term stability of the
  {K}epler-60 exoplanetary system, Monthly Notices of the Royal Astronomical
  Society 509 (2022) 884.

\bibitem{HHH2022}
Z.~Huang, G.~Huang, A.~Hu, Application of explicit symplectic integrators in a
  magnetized deformed {S}chwarzschild black spacetime, The Astrophysical
  Journal 925 (2022) 158.

\bibitem{G2021}
B.~Ghanbari, On detecting chaos in a prey-predator model with prey’s
  counter-attack on juvenile predators, Chaos Solitons \& Fractals 150 (2021)
  111136.

\bibitem{SAZ2021}
A.~F. Steklain, A.~Al-Ghamdi, E.~E. Zotos, Using chaos indicators to determine
  vaccine influence on epidemic stabilization, Physical Review E 103 (2021)
  032212.

\bibitem{BRM2024}
E.~Blumenthal, J.~W. Rocks, P.~Mehta, Phase transition to chaos in complex
  ecosystems with nonreciprocal species-resource interactions, Physical Review
  Letters 132 (2024) 127401.

\bibitem{SG2010}
C.~Skokos, E.~Gerlach, Numerical integration of variational equations, Physical
  Review E 82 (2010) 036704.

\bibitem{GES2012}
E.~Gerlach, S.~Eggl, C.~Skokos, Efficient integration of the variational
  equations of multidimensional {H}amiltonian systems: {A}pplication to the
  {F}ermi–{P}asta–{U}lam lattice, International Journal of Bifurcation and
  Chaos 22 (2012) 1250216.

\bibitem{SS2018}
B.~Senyange, C.~Skokos, Computational efficiency of symplectic integration
  schemes: application to multidimensional disordered {K}lein--{G}ordon
  lattices, The European Physical Journal Special Topics 227 (2018) 625.

\bibitem{TSR2001}
G.~Tancredi, A.~S{\'{a}}nchez, F.~Roig, A comparison between methods to compute
  {L}yapunov exponents, The Astronomical Journal 121 (2001) 1171.

\bibitem{MH2018}
L.~Mei, L.~Huang, Reliability of {L}yapunov characteristic exponents computed
  by the two-particle method, Computer Physics Communications 224 (2018) 108.

\bibitem{HZNKWS2022}
M.~Hillebrand, S.~Zimper, A.~Ngapasare, M.~Katsanikas, S.~Wiggins, C.~Skokos,
  Quantifying chaos using {Lagrangian} descriptors, Chaos: An Interdisciplinary
  Journal of Nonlinear Science 32 (2022) 123122.

\bibitem{BGM2023}
A.~Bazzani, M.~Giovannozzi, C.~E. Montanari, G.~Turchetti, Performance analysis
  of indicators of chaos for nonlinear dynamical systems, Phys. Rev. E 107
  (2023) 064209.

\bibitem{DK2018}
M.-F. Danca, N.~Kuznetsov, Matlab code for {L}yapunov exponents of
  fractional-order systems, International Journal of Bifurcation and Chaos 28
  (2018) 1850067.

\bibitem{J1987}
K.~L. Johnson, Contact mechanics, Cambridge university press, 1987.

\bibitem{LWKH2008}
C.~Lee, X.~Wei, J.~W. Kysar, J.~Hone, Measurement of the elastic properties and
  intrinsic strength of monolayer graphene, Science 321 (2008) 385.

\bibitem{CPGC2009}
E.~Cadelano, P.~L. Palla, S.~Giordano, L.~Colombo, Nonlinear elasticity of
  monolayer graphene, Phys. Rev. Lett. 102 (2009) 235502.

\bibitem{Schlegel2003}
H.~Schlegel, Exploring potential energy surfaces for chemical reactions: an
  overview of some practical methods, Journal of Computation Chemistry 24
  (2003) 1514.

\bibitem{SW2018}
W.~R. Smith, W.~Qi, Molecular simulation of chemical reaction equilibrium by
  computationally efficient free energy minimization, ACS Central Science 4
  (2018) 1185.

\bibitem{Naidoo2021}
K.~J. Naidoo, T.~Bruce-Chwatt, T.~Senapathi, M.~Hillebrand, Multidimensional
  free energy and accelerated quantum library methods provide a gateway to
  glycoenzyme conformational, electronic, and reaction mechanisms, Accounts of
  Chemical Research 54 (2021) 4120.

\bibitem{HMKGS2020}
M.~Hillebrand, B.~Many~Manda, G.~Kalosakas, E.~Gerlach, C.~Skokos, Chaotic
  dynamics of graphene and graphene nanoribbons, Chaos 30 (2020) 063150.

\bibitem{HH1964}
M.~H{\'e}non, C.~Heiles, The applicability of the third integral of motion:
  some numerical experiments, The Astronomical Journal 69 (1964) 73.

\bibitem{FORD1992}
J.~Ford, The {F}ermi-{P}asta-{U}lam problem: {P}aradox turns discovery, Physics
  Reports 213 (1992) 271.

\bibitem{BI2005}
G.~P. Berman, F.~M. Izrailev, The {F}ermi-{P}asta-{U}lam problem: Fifty years
  of progress, Chaos 15 (2005) 015104.

\bibitem{SPM2020}
W.~Szumi{\'n}ski, M.~Przybylska, A.~J. Maciejewski, Comment on {“Hyperchaos
  in constrained Hamiltonian system and its control” by J. Li, H. Wu and F.
  Mei}, Nonlinear Dynamics 101~(1) (2020) 639--654.

\bibitem{HSP2019}
S.~He, K.~Sun, Y.~Peng, Detecting chaos in fractional-order nonlinear systems
  using the smaller alignment index, Physics Letters A 383 (2019) 2267.

\bibitem{Ma2016}
D.-Z. Ma, Z.-C. Long, Y.~Zhu, Application of indicators for chaos in chaotic
  circuit systems, International Journal of Bifurcation and Chaos 26 (2016)
  1650182.

\bibitem{Chater2022}
W.~Chatar, M.~El~Ghamari, J.~Kharbach, M.~Benkhali, R.~Masrour, A.~Rezzouk,
  M.~Ouazzani-Jamil, Detecting order and chaos by the m{LE}, {SALI} and {GALI}
  methods in three-dimensional nonlinear {Y}ang–{M}ills system, International
  Journal of Bifurcation and Chaos 32 (2022) 2250145.

\bibitem{PMDN2004}
T.~Mai, A.~Dhar, O.~Narayan, Equilibration and universal heat conduction in
  {F}ermi-{P}asta-{U}lam chains, Physical Review Lett. 98 (2007) 184301.

\bibitem{OVPL2015}
M.~Onorato, L.~Vozella, D.~Proment, Y.~V. Lvov, Route to thermalization in the
  $\alpha$-{F}ermi-{P}asta-{U}lam system, Proceedings of the National Academy
  of Sciences 112 (2015) 4208.

\bibitem{DCF2017}
C.~Danieli, D.~K. Campbell, S.~Flach, Intermittent many-body dynamics at
  equilibrium, Physical Review E 95 (2017) 060202.

\bibitem{ZWZZ2005}
H.~Zhao, Z.~Wen, Y.~Zhang, D.~Zheng, Dynamics of solitary wave scattering in
  the {F}ermi-{P}asta-{U}lam model, Physical Review Letters 94 (2005) 025507.

\bibitem{FG2008}
S.~Flach, A.~V. Gorbach, Discrete breathers—{A}dvances in theory and
  applications, Physics Reports 467 (2008) 1.

\bibitem{CDRT1998}
T.~Cretegny, T.~Dauxois, S.~Ruffo, A.~Torcini, Localization and equipartition
  of energy in the $\beta$-{FPU} chain: {C}haotic breathers, Physica D 121
  (1998) 109.

\bibitem{YR2024}
H.~Yan, M.~Robnik, Chaos and quantization of the three-particle generic
  {F}ermi-{P}asta-{U}lam-{T}singou model. {I}. {D}ensity of states and spectral
  statistics, Physical Review E 109 (2024) 054210.

\bibitem{ABS06}
C.~Antonopoulos, T.~Bountis, C.~Skokos, Chaotic dynamics of {N}-degree of
  freedom {H}amiltonian systems, International Journal of Bifurcation and Chaos
  16 (2006) 1777--1793.

\bibitem{AB2006}
C.~Antonopoulos, T.~Bountis, Stability of simple periodic orbits and chaos in a
  {F}ermi-{P}asta-{U}lam lattice, Physical Review E 73 (2006) 056206.

\bibitem{D2004}
T.~Dauxois, A.~Litvak-Hinenzon, R.~MacKay, A.~Spanoudaki, Energy localisation
  and transfer, World Scientific, 2004.

\bibitem{MMS2020}
H.~Moges, T.~Manos, C.~Skokos, On the behavior of the generalized alignment
  index ({GALI}) method for regular motion in multidimensional {H}amiltonian
  systems, Nonlinear Phenomena in Complex Systems 123 (2020) 153.

\bibitem{SBA2008}
C.~Skokos, T.~Bountis, C.~Antonopoulos, Detecting chaos, determining the
  dimensions of tori and predicting slow diffusion in {F}ermi--{P}asta--{U}lam
  lattices by the generalized alignment index method, The European Physical
  Journal Special Topics 165 (2008) 5.

\bibitem{BCFLMM2013}
S.~Blanes, F.~Casas, A.~Farres, J.~Laskar, J.~Makazaga, A.~Murua, New families
  of symplectic splitting methods for numerical integration in dynamical
  astronomy, Applied Numerical Mathematics 68 (2013) 58.

\bibitem{FLBCMM2013}
A.~Farr{\'e}s, J.~Laskar, S.~Blanes, F.~Casas, J.~Makazaga, A.~Murua, High
  precision symplectic integrators for the solar system, Celestial Mechanics
  and Dynamical Astronomy 116 (2013) 141.

\bibitem{DMMS2019}
C.~Danieli, B.~Many~Manda, T.~Mithun, C.~Skokos, Computational efficiency of
  numerical integration methods for the tangent dynamics of many-body
  {H}amiltonian systems in one and two spatial dimensions, Mathematics in
  Engineering 1 (2019) 447.

\bibitem{FJZWL2020}
H.~Fan, J.~Jiang, C.~Zhang, X.~Wang, Y.-C. Lai, Long-term prediction of chaotic
  systems with machine learning, Phys. Rev. Res. 2 (2020) 012080.

\bibitem{PSM2023}
M.~Przybylska, W.~Szumi{\'n}ski, A.~J. Maciejewski, {Destructive relativity},
  Chaos: An Interdisciplinary Journal of Nonlinear Science 33 (2023) 063156.

\bibitem{SPM2024}
W.~Szumi{\'n}ski, M.~Przybylska, A.~J. Maciejewski, Chaos and integrability of
  relativistic homogeneous potentials in curved space, Nonlinear Dynamics 112
  (2024) 4879.

\end{thebibliography}






\end{document}